\documentclass{jfm}
\usepackage{graphicx}
\usepackage{natbib}
\usepackage{setspace}
\usepackage{caption} 
\usepackage{subcaption} 
\usepackage{placeins}
\usepackage{float, lscape}
\usepackage[dvipsnames]{xcolor}
\usepackage{sidecap}
\usepackage{amssymb}
\usepackage{mathrsfs}
\usepackage{siunitx}
\usepackage{enumitem}
\usepackage{tikz}
\usepackage{float}

\title[Dynamics of cavity collapse]{Dynamics of collapse of free-surface  bubbles: effects of gravity and viscosity} 

\author[Sangeeth, K., Puthenveettil, B. A., and Hopfinger, E. J.]{Sangeeth Krishnan$^1$\thanks{Email address for correspondence: sangeeth.krishnan@icts.res.in},\ns Baburaj~A.~Puthenveettil$^2$and~E.~J.~Hopfinger$^3$ \break}

\affiliation{$^1$International Centre for Theoretical Sciences, Tata Institute of Fundamental Research, Bengaluru-560 089, India\\[\affilskip]$^2$Department of Applied Mechanics, Indian Institute of Technology Madras, Chennai- 600 036, India\\[\affilskip]$^2$LEGI-CNRS, Université Grenoble Alpes, CS40700, 38052, Grenoble, France}

\pubyear{} \volume{} \pagerange{}
\date{?; revised ?; accepted ?. - To be entered by editorial office}
\begin{document} 
\maketitle
\begin{abstract} 
The rupture of the thin film at the top of a  bubble floating at a liquid-gas interface  leads to the axisymmetric collapse of the bubble cavity. We present scaling laws for such a cavity collapse,  established from experiments conducted with bubbles spanning a wide range of Bond (${10^{-3}<Bo\leq1}$) and Ohnesorge numbers (${10^{-3}<Oh<10^{-1}}$), defined with the bubble radius $R$. The cavity collapse is a capillary-driven process, with a dependency on viscosity and gravity affecting, respectively, precursory capillary waves on the cavity boundary, and the static bubble shape. The collapse is characterised by tangential and normal velocities of the kink, formed by the intersection of the concave cavity opening after the top thin film rupture, with the convex bubble cavity boundary. The tangential velocity $U_t$ is constant during the collapse and is shown to be  $U_t=4.5~U_c{\mathcal{W}}_R$, where $U_c$ is the capillary velocity and  ${\mathcal{W}}_R(Oh,Bo)={(1-\sqrt{Oh {\mathscr{L}}} )^{-1/2}}$  is the wave resistance factor due to the precursory capillary waves,    with  $\mathscr{L}(Bo)$ being the path correction of the kink motion. The movement of the kink in the normal direction is part of the inward shrinkage of the whole cavity due to the sudden  reduction of gas pressure inside the bubble cavity after the thin film rupture. This normal velocity is shown to scale as $U_c$ in the equatorial plane, while at the bottom of the cavity $\overline{U}_{nb}=U_c(Z_c/R)({\mathcal{W}_R}/ {\mathscr{L}})$, where $Z_c(Bo)$ is the static cavity depth. The total volume flux of cavity-filling, which is entirely contributed by this shrinking, scales as ${Q_T\simeq2\pi R Z_c U_c}$; remains a constant throughout the collapse.
\end{abstract}

\section{ Introduction}\label{introduction}
\begin{figure}
    \centering
    \begin{subfigure}{0.24\textwidth}
        \includegraphics[width=\textwidth]{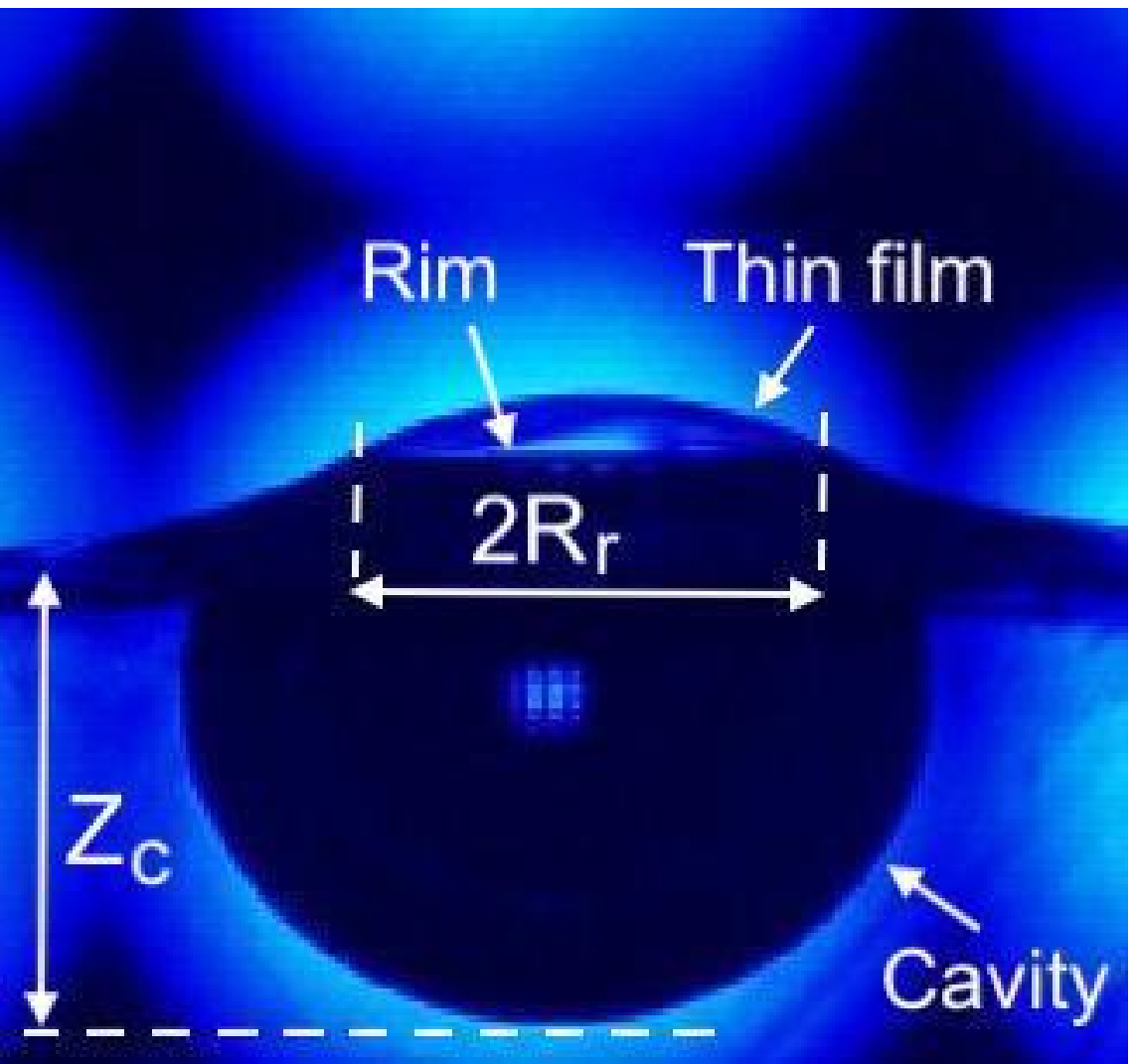}
        \caption{}
        \label{B6}
    \end{subfigure}
    \begin{subfigure}{0.24\textwidth}
        \includegraphics[width=\textwidth]{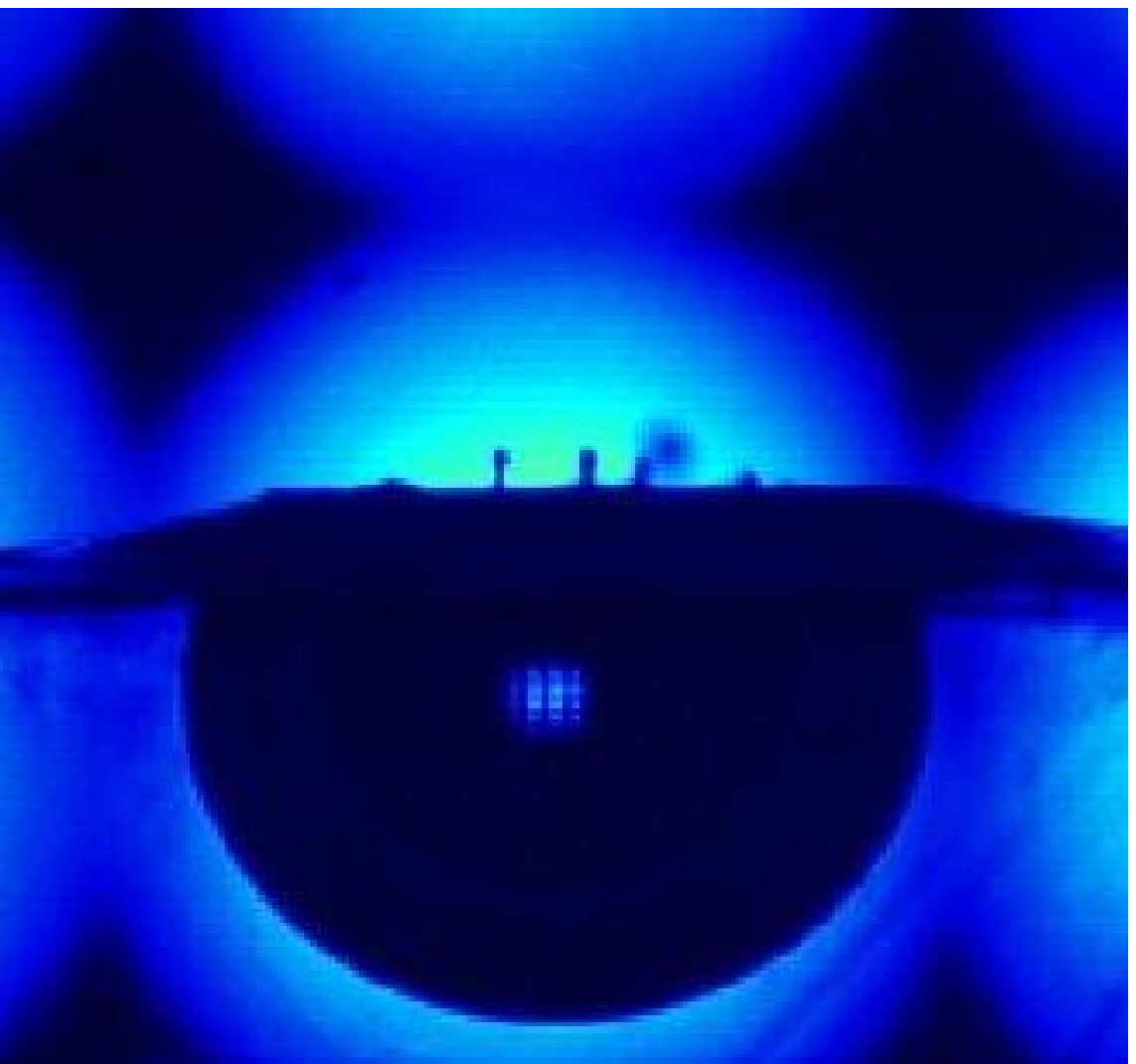}
        \caption{}
        \label{B9}
    \end{subfigure}
    \begin{subfigure}{0.24\textwidth}
        \includegraphics[width=\textwidth]{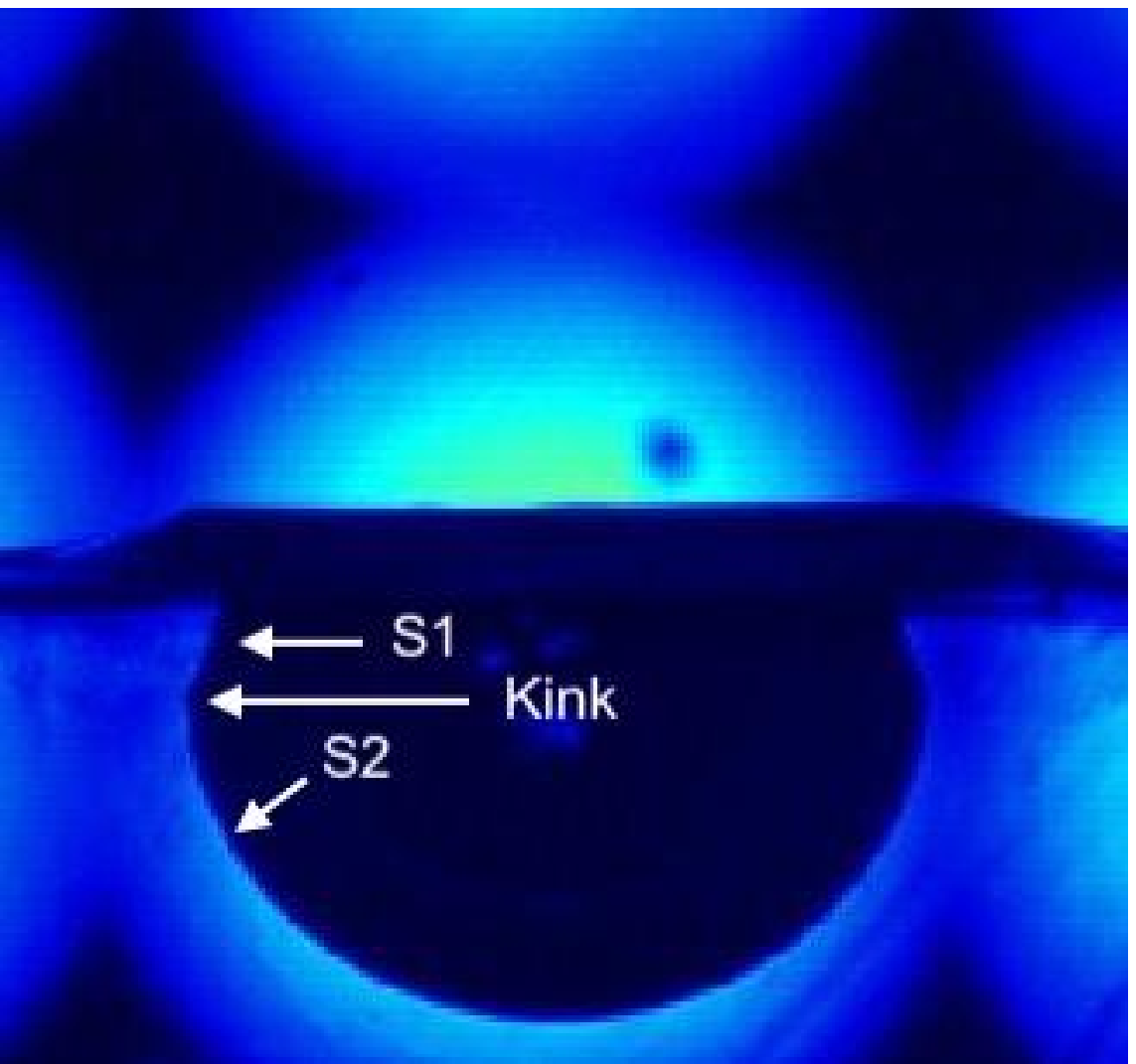}
        \caption{}
        \label{B12}
    \end{subfigure}
		\begin{subfigure}{0.24\textwidth}
        \includegraphics[width=\textwidth]{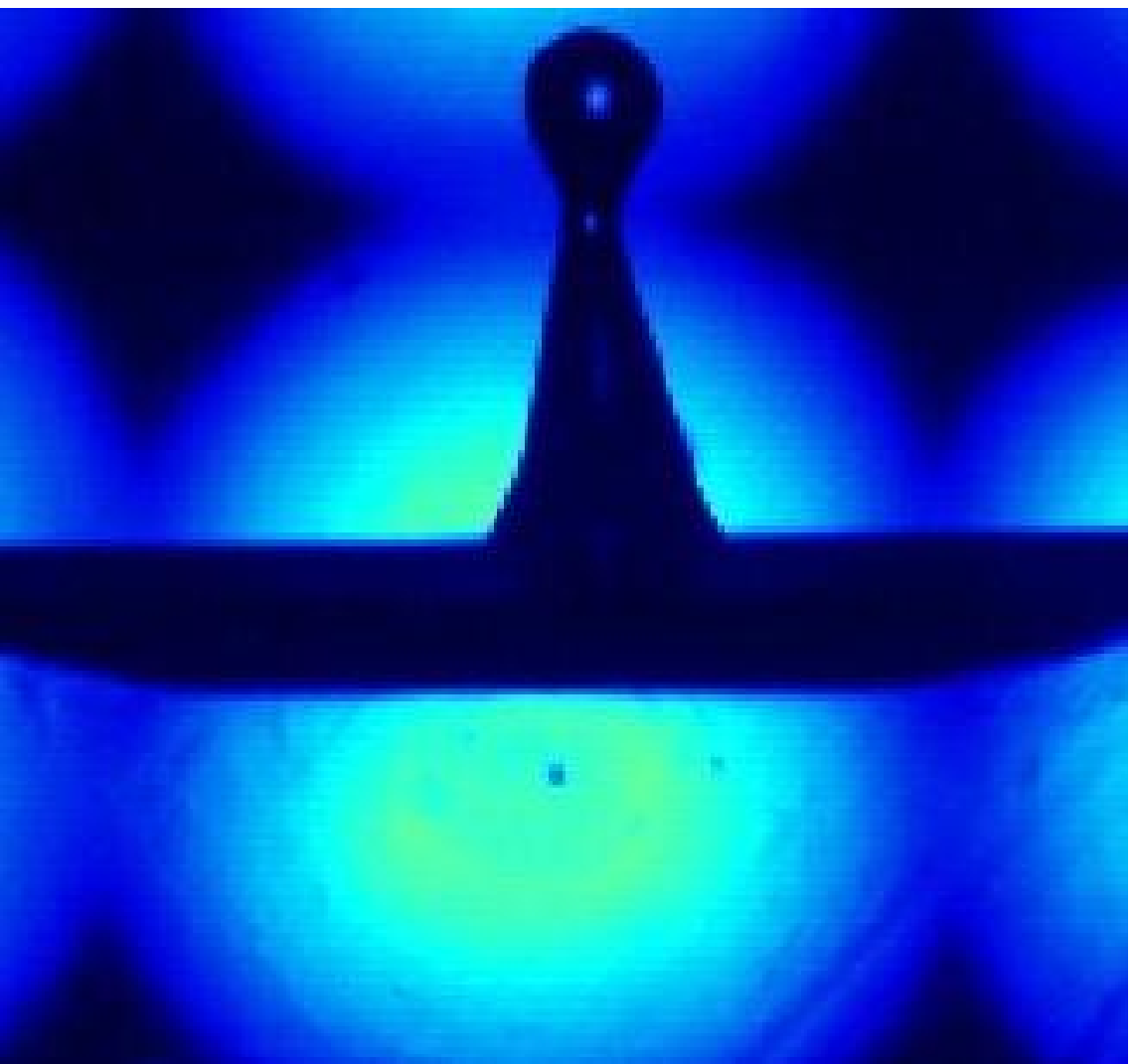}
        \caption{}
        \label{Bd}
    \end{subfigure}
	
				    \caption{Stages of bubble collapse at the free-surface. (a),  static bubble in water with a radius $R~\approx~2$mm; (b),  rupturing of thin film; (c), beginning of the cavity collapse; (d), jet at the free-surface (d). The time gaps between the images are of the order of $10^{-4}$s.}
    \label{introbubbleburst}
\end{figure}

 A bubble at a liquid-gas interface is characterised by a cavity, capped from above by a spherical thin film, and joined at a circular rim, as shown in figure~\ref{B6}. The rupture of the thin film leaves an unstable cavity at the interface, which collapses axisymmetrically and generates a high-velocity jet \citep{woodcock1953giant,kientzler1954photographic}; figure~\ref{introbubbleburst} shows an image sequence of a  bubble bursting at the water-air interface.  The bursting of these free-surface bubbles is an important transport mechanism in many applications. Mass transport from the liquid surfaces to the ambient air is of importance in air-sea exchange and the spread of pathogens  \citep{blanchard1963electrification,MacIntyre1,spiel1995births,Walls2015_drops,joung2017bioaerosol,Sampath_GeoPhyRes2019,yang2023enhanced}. Bubble bursting has also been investigated in connection with the reverse mass transport observed in the mixing of the oil spill in the ocean \citep{feng2014nanoemulsions}, and in the context of the creation of intense stress zones in bioreactors \citep{BSB,walls2017quantifying}.

Most of these studies on free-surface bubble collapse have focused on the dynamics of jetting.  The general consensus on jetting is that  when Ohnesorge numbers,   $Oh=\mu/\sqrt{\sigma\rho R}<0.037$, the jet velocity $U_j$ scales with the capillary velocity $U_c=\sqrt{\sigma/\rho R}$ (see Table~\ref{fluid_properties2} for definitions of symbols),  provided, the bubbles are small, such that Bond numbers,  $Bo=\rho g R^2/\sigma <0.1$. However, for larger bubbles, when $Bo>0.1$,  the jet velocity can deviate substantially from the capillary velocity $U_c$ due to gravity effects   
\citep{krishnan_hopfinger_puthenveettil_2017,DeikePRF2018,AlfonsoPRL2017,Alfonso_2018_PhysRevFluids,gordillo2019capillary}.  A jet Weber number scaling,  $We_j=\rho{U_j}^2 R/\sigma\sim (Z_c/R)^2$,    proposed by \cite{krishnan_hopfinger_puthenveettil_2017}, explains the effect of gravity on jet velocity through the static depth of the bubble cavity $Z_c$ (see figure~\ref{introbubbleburst}(a)), where $Z_c$ is a function of Bond number. The scaling of the velocity and size of the jet ejected from the bursting free-surface bubble, or the velocity and size of the drop due to the break up of such a jet, is still not fully understood.

The velocity and size of the jet/drop that is ejected is inherently related to the dynamics of the collapse of the cavity created by the bubble at the free-surface, which has not been studied well. It is known that the velocity of the collapsing cavity is an order of magnitude less than the jet velocity \citep{Sangeeth2015207}. As seen in the images of surface bubble cavity collapse in figure~\ref{introbubbleburst}, after the thin surface film rupture, the hole expansion  creates a concave boundary  (as seen from the liquid side) S1 in figure~\ref{B12} \citep{sangeeth_hole_expn}, with the formation of a kink at its intersection with the convex cavity shape S2 in figure~\ref{B12}. The kink moves tangentially along the boundary with a velocity $U_t$, while at the same time, the cavity shrinks with a velocity normal to the boundary $U_n$ due to the excess capillary pressure, after the gas pressure drops when the cavity opens. Capillary waves,  similar to the waves observed earlier in   steep gravity waves \citep{perlin1993} and  Faraday waves \citep{Das_Hopfinger_Faradaywaves2008}, move ahead of the kink.   The reduction of the amplitude of these  precursory capillary waves  is  proportional  to   $Oh^{1/2}$, which is valid till the complete suppression of the waves at ${Oh\approx 0.02}$  \citep{krishnan_hopfinger_puthenveettil_2017,gordillo2019capillary}. Such progressive viscous damping of these waves  results   in an increase in the jet velocity  \citep{ghabache2014physics}.  

While the stages of collapse described above, and shown in figure~\ref{introbubbleburst}, have been well identified \citep{MacIntyre1,duchemin2002jet,san2011size,Brasz_PhysRevFluids2018}, quantitative information on the velocities of collapse and the related mass fluxes are not available; neither are any scaling laws for these available. The two-dimensionality of the moving kink, and the lack of top-down symmetry of the interface during flow-focusing prevent the use of one-dimensional, Rayleigh-Plesset equation based models, often used to study the cavities at the free-surface formed due to impacting objects \citep{oguz1993dynamics,burton2005scaling,Bartolo_dropimpact_06PRL,Bergmann2006PRL,duclaux2007}. Even though it has been found that the kink moves with a constant velocity proportional to the capillary velocity scale \citep{Sangeeth2015207,krishnan_hopfinger_puthenveettil_2017,gordillo2019capillary}, since precursory capillary waves occur, and the total path length depends on $Bo$, it is not clear whether this velocity is also dependent on $Oh$ and $Bo$. Though the effect of precursory capillary waves on jet velocity has been studied previously \citep{ghabache2014physics,krishnan_hopfinger_puthenveettil_2017,AlfonsoPRL2017,DeikePRF2018,Alfonso_2018_PhysRevFluids,Ganan-calvo_lopez-herrera_2021,gordillo2019capillary,blanco-rodriguez_gordillo_2021}, the effect of these waves on the collapsing cavity surface has not been addressed.

Most other studies \citep{AlfonsoPRL2017,Alfonso_2018_PhysRevFluids,Ganan-calvo_lopez-herrera_2021,Ismail2018Softmatter,Universal_bubblejetcavity_Eggers_deike,blanco-rodriguez_gordillo_2021} limit their focus on the dynamics of flow-focusing in a small region at the cavity bottom, where viscosity also dictates the length scale. \cite{Ganan-calvo_lopez-herrera_2021}  proposed the spherically averaged velocity during the flow-focusing to scale as $W\sim(V_{\mu}/Oh_L)\psi(Oh,Oh_L,Bo)$, where $V_{\mu}=\sigma/\mu$, and $Oh_L$ is the Ohnersorge number based on the length scale at the bottom of the cavity at flow focusing;  in the limit $Oh\ll0.04$, $W$ then tends to the capillary velocity $U_c$ \citep{AlfonsoPRL2017,Alfonso_2018_PhysRevFluids}. \cite{gordillo2019capillary} and \cite{blanco-rodriguez_gordillo_2021}   assumed a purely horizontal and radially inward flow during the flow-focusing at the cavity bottom and modelled the flow using a vertical array of sinks placed along meridional centre line, with the length of the array decided by the size of the bubble and the wavelength of the capillary waves moving ahead of the kink. However, as we mentioned above, and discuss in detail later, the flow focusing region has both radial (spherical) and tangential velocities, which actually scale differently. \cite{Universal_bubblejetcavity_Eggers_deike}  showed that the shapes of the collapsing cavity to be self-similar, showing a $|t_s-t|^{2/3}$  scaling, where $t_s-t$ is the time to the singularity, with $t_s$ being the instant of fluid convergence at the cavity bottom, similar to the scaling of \cite{zeff2000singularity}  in Faraday wave collapse; such self similarity is however present only for $0.014<~Oh~\leq~0.04$  at small Bond numbers ($Bo\ll 0.1$), when precursory capillary waves are absent.

In the present paper we present detailed experiments to study the dynamics of the cavity in a free-surface bubble collapse, by analysing which, we obtain scaling laws for the duration of collapse, the various velocities of collapse and the volume fluxes involved in the collapse. We show that the precursory capillary waves reduce the velocity of the moving kink in the tangential and the normal directions. The volume fluxes are entirely due to shrinkage of the cavity walls in the normal direction with a direct dependency on the cavity depth $Z_c$. The effects of viscosity and gravity on cavity collapse can be quantified using three parameters: the  path correction $\mathscr{L}(Bo)$, the wave resistance factor ${\mathcal{W}}_R(Oh,Bo)$ and the aspect ratio of the cavity   ${Z_c(Bo)}/{R}$. These aspects of cavity collapse are essential for the understanding of the effects of viscosity and gravity on jetting.
The paper is organised as follows. In \S~\ref{expsetup}  the experimental setup and conditions are presented. Then, in \S~\ref{bubble_collapse}, different aspects of cavity collapse, namely the velocities in tangential and normal directions of the collapsing cavity wall, as well as the total time of cavity collapse,  are discussed. Scaling relations are established that explain  the effect of gravity and viscosity on these parameters.  In \S~\ref{massfilltext} the volume influxes corresponding to the cavity collapse velocities are determined, before concluding in \S~\ref{conclusions}.   

\section{Experimental conditions}
\label{expsetup}
The experiments were conducted in two transparent containers of   cross-sectional areas of 5 $\times$ 5 cm$^2$ and 3.5 $\times$ 5 cm$^2$,  filled with various fluids, \emph{viz.}, distilled water, various glycerol-water mixtures with  weight of glycerine of  48\%, 55\%, 68\%, and 72\%, (hereinafter referred to as GW48, GW55, GW68, and GW72), ethanol and 2-propanol. Table~\ref{fluid_properties2} shows the properties of these  fluids. In order to avoid meniscus effects, the containers were filled with the desired liquids up to the edge. 
\begin{table}
  \begin{center}
	\def~{\hphantom{0}}
  \begin{tabular}{l*{9}{c}r}
     &$\mu$ & $\rho$& $\sigma$  &    $R$ &   $Bo$ & $Oh$ & $t_c$ & $U_c$  \\[3pt]
      &mPas  & kgm$^{-3}$& Nm$^{-1}$  &    mm &   &  & ms & ms$^{-1}$  \\[3pt]
      Water & 1.005        & 1000 & 0.072                 & 0.175-4.1                          & 0.004-2.27     & 0.0019-0.009 &0.3-30.7          & 0.64-0.13\\
			Ethanol & 1.144        & 789  & 0.022                &  0.19-1.16                            & 0.013-0.47     & 0.008-0.02 &0.5-7.5          & 0.38-0.16\\
       2-proponol & 2.073        & 781   & 0.018                & 1.46-2.41                            & 0.9-2.4     & 0.011-0.014 &11.6-24.6          & 0.13-0.1\\
			GW48(30$^{\circ}$C) & 3.9        & 1115    & 0.068           & 0.42-3.4                           & 0.029-1.9     & 0.0076-0.021 &1.1-25.4          & 0.38-0.13\\
			GW48(20$^{\circ}$C) & 5.5        & 1120     & 0.068             &  0.81-1.96                          & 0.1-0.62     & 0.014-0.022 &3-11.1          & 0.27-0.18\\
			GW55 & 8        & 1140     & 0.067             & 0.71-2.3                         & 0.08-0.88     & 0.019-0.035 &2.5-14.4          & 0.29-0.16\\
			GW 68 & 12.414        & 1170    & 0.066           & 0.48-2.3                         & 0.04-0.89     & 0.03-0.064 &1.4-14.7          & 0.34-0.16\\
			GW 72 & 16.616        & 1181    & 0.064           & 0.6-3.6                          & 0.063-2.4     & 0.032-0.079 &2-29.3         & 0.3-0.12\\
  \end{tabular}
  \caption{The properties of the fluids used in the  experiments and the corresponding  dimensionless parameters. The fluid properties $\sigma$, $\rho$ and $\mu$ are the surface tension, density and viscosity, respectively. $g$ is   the acceleration due to gravity. The  Bond number  ${Bo=\rho g R^2/\sigma}$, the Ohnesorge number   $Oh=\mu/\sqrt{\sigma \rho R}$, the capillary time scale $t_c =\sqrt{\rho R^{3}/\sigma}$ and the capillary velocity scale  $U_c=\sqrt{\sigma/\rho R}$.}
  \label{fluid_properties2}
	  \end{center}
\end{table}
Fine capillaries of various sizes, connected to a constant discharge syringe pump, were kept immersed in the working fluid to create bubbles of different, equivalent, spherical radii $R$.  Low discharge rates were maintained, so that the bubbles were in the periodic discharge regime  \citep{oguz1993dynamics}. To prevent   variation in bubble sizes from each capillary, the orientations of the capillaries were maintained the same throughout the experiments \citep{Doshi_persistence_universality}.     The bubble  occupied the centre of the free-surface. We used La Vision ProHS  ({frame~rate~$\leq$~19000Hz}) and Photron SA4  ({frame~rate~$\leq$~100000Hz}) cameras for high-speed imaging of the side views of the dynamics of the cavities. A high-intensity green LED array was used for back lighting. The image acquisition rates  met the condition that $t_i < 1/\left| dU_{abs}/ds \right|$, where {$t_i$=1/(frame~rate)} and $dU_{abs}/ds$ is the  spatial gradient of the absolute velocity of the kink along the cavity. The spatial resolution was such that $\Delta Z_i < U_{abs}~t_{exp}$, where $\Delta Z_i$ is the size of each pixel and $t_{exp}$ is the exposure time. The lowest and the highest resolutions for the imaging were 27$\mu$m/pix and 3.4$\mu$m/pix, respectively. 

\begin{figure}
 \centering
    \begin{subfigure}{0.39\textwidth}
        \includegraphics[width=\textwidth]{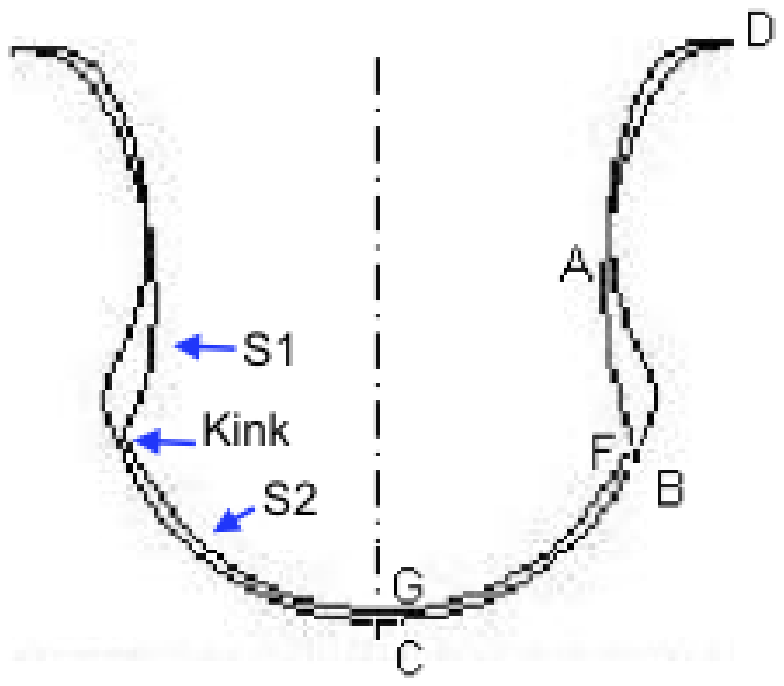}
        \caption{}
        \label{QsQt_expt}
    \end{subfigure}
\centering
 \begin{subfigure}{0.37\textwidth}
        \includegraphics[width=\textwidth]{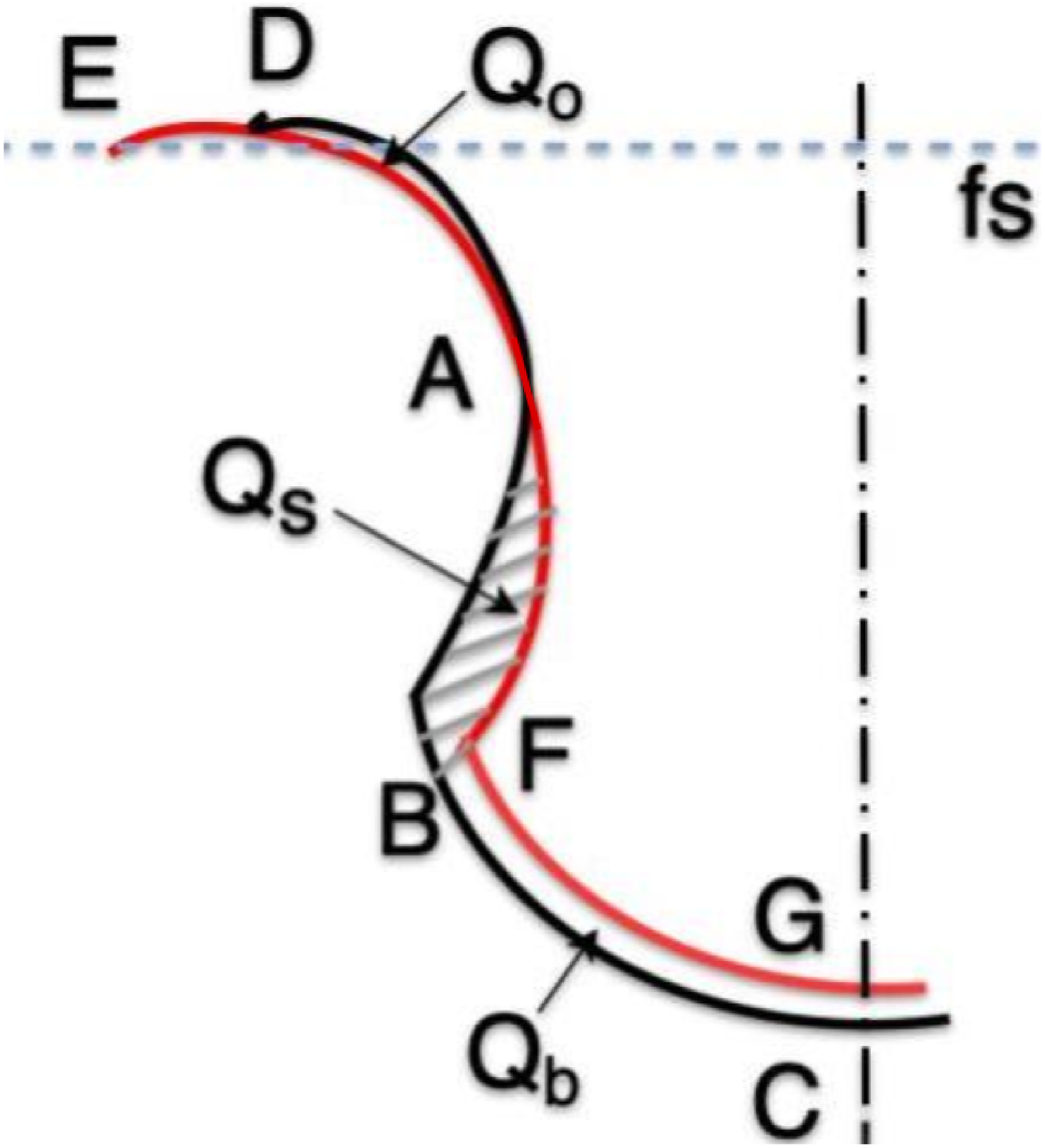}
        \caption{}
        \label{QtQn_labelled}
    \end{subfigure}
    \centering
    \begin{subfigure}{0.49\textwidth}
        \includegraphics[width=\textwidth]{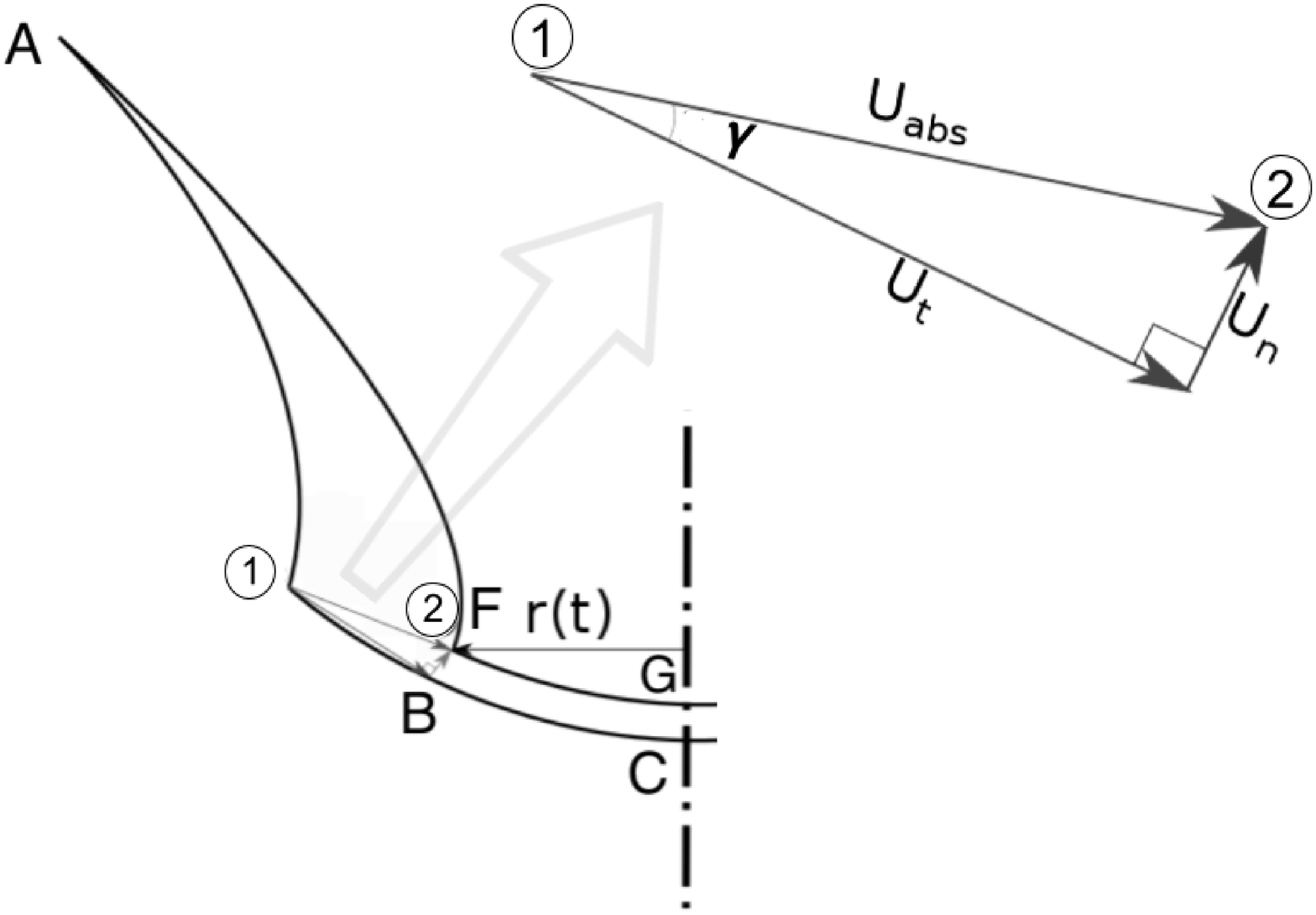}
        \caption{}
        \label{cavitycollapse_labelled}
    \end{subfigure}
     \centering
    \begin{subfigure}{0.48\textwidth}
        \includegraphics[width=\textwidth]{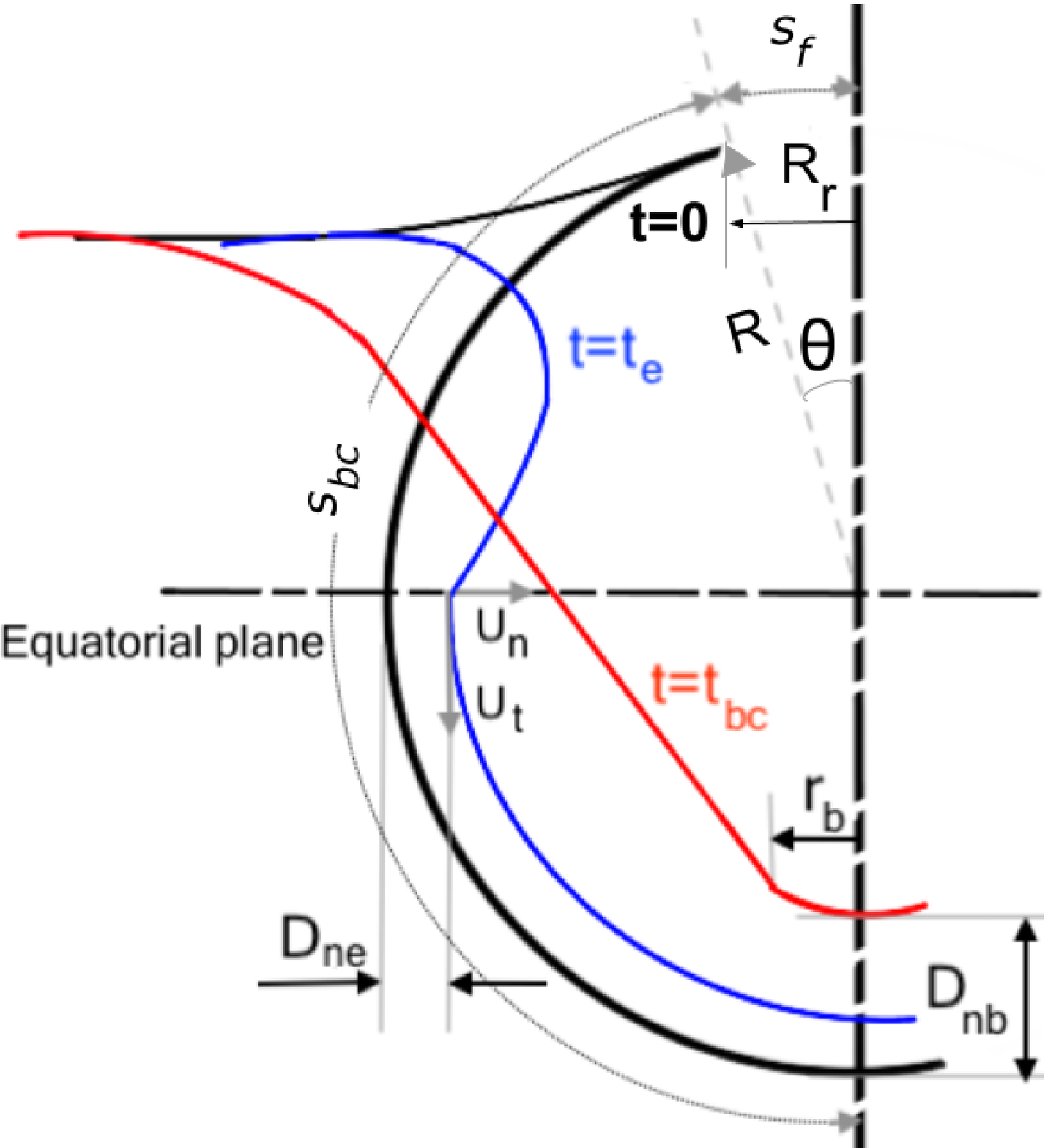}
        \caption{}
        \label{DeDnb_labelled}
    \end{subfigure}
        \caption{Schematics depicting the parameters and terminologies describing the cavity collapse. (a), the actual cavity contours at two time instances, extracted from experiments with a bubble of $R=0.175$mm in water showing the kink. S1 and S2 denote the concave and the convex boundaries of the cavity (also see figure~\ref{img188}). (b), the side volume flux $Q_s$ and the bottom volume flux $Q_b$ due to the difference between the side and the bottom cavity contours at two successive time instances. Similarly,  $Q_o$ is the volume out-flux at the top, estimated as the difference between the cavity contours at the free-surface fs.  (c), schematic of the collapsing cavity contours at two time instances, with the position of the kink at the different times marked as \textcircled{1} and \textcircled{2}.   The  absolute, tangential and normal velocities of the kink are shown in the associated vector triangle. (d), schematic of the  cavity contours at the following times.  (i) $t=0$ (black); the initial cavity contour when the thin film is ruptured. (ii) $t=t_e$ (blue); the cavity contour when the kink has arrived at the equatorial plane of the cavity   (denoted by the horizontal line)   showing the equatorial cavity shrinkage $D_{ne}$. (iii) $t=t_{bc}$ (red); the cavity contour when the kink has arrived at the bottom when the cavity has the form of a truncated cone with bottom radius $r_b$ and vertical cavity retraction $D_{nb}$.} 
    \label{}
\end{figure}

The following length measurements were done by counting the pixels between the appropriate liquid-gas interfaces seen in the images. The  equivalent spherical bubble radii ($R$) were measured from the images of the rising bubbles generated at the capillaries.  The  cavity shrinking lengths along the equatorial plane $D_{ne}$  and  along the  vertical plane $D_{nb}$, as well as the bottom radius of the conical cavity $r_b$  (see figure~\ref{DeDnb_labelled}), were measured as a function of time from the instantaneous images of the collapsing cavity.
The time corresponding to each image was estimated from the frame rate of recording,  with the zero time being the time of thin film rupture. The total time of cavity collapse $t_{bc}$ was measured by counting the number of images starting from the thin film rupture till the  cavity becomes conical  (see figure~\ref{DeDnb_labelled}). The times  corresponding to the lengths $D_{ne}$ and $D_{nb}$ were  measured similarly.

 Velocities of the moving kink, in directions tangential and normal to the cavity surface ($U_t$ and $U_n$), were measured by resolving the absolute velocities of the kink $U_{abs}$  in two mutually orthogonal directions, as shown in figure~\ref{cavitycollapse_labelled}.  The absolute displacement of the kink was measured by finding its coordinates at subsequent instances, with $U_{abs}$ being obtained  by dividing the absolute displacement by the time gap between the images. The angle $\gamma$ (see figure~\ref{cavitycollapse_labelled}) was measured throughout the collapse duration by finding $\tan \gamma$ by vectorial decomposition of the absolute velocity along the tangential and the normal directions. Polynomial fits of the progressive displacements in tangential direction ($d_t$)   as a function of time, similar to that shown in the inset (a) of figure~\ref{Vt_inst1} were used to calculate   $U_t(t)$  by taking the time derivative of the  fits. In the same way, the normal velocity $U_n$ was estimated from progressive normal displacements.
 
 We define three volume fluxes related with the cavity boundary movement: the side (tangential) volume influx $Q_s$, the bottom (normal) influx $Q_b$ and the side volume out-flux $Q_o$, with the total filling rate being $Q_{T}=Q_s+Q_b$. The area ABF shown in figure~\ref{QtQn_labelled} is the  area swept by two successive positions of the kink as it travels {along the cavity surface and} inwards,  with the corresponding side (tangential) volume influx being $Q_s$. Similarly, the area BCGF is the area swept by the normal motion of the bottom regions of the cavity, with corresponding  volume influx being $Q_b$.   The volume out-flux $Q_o$, corresponding to the area DA, was only measured for a single bubble.  These volumes were measured as follows: the edges of  the collapsing cavity were extracted from images using  Canny or Sobel edge detection criteria, depending on the noise levels in the image sequence.  Two successive contours  were superimposed to produce a sequence of edge pairs (see figure~\ref{QsQt_expt}) with time. Within two successive contours, the radial distance ($r_p$) of each pixel  and the total number of pixels $\eta_{p}$ were measured at each time. The volume contributed by a square pixel inside the two edges, $2 \pi r_p  {\Delta Z_i}^2$, was estimated. This process was repeated for all the pixels inside the contours,  and the volume contributions from each pixel were added. The value of this cumulative volume was then  divided by the time gap between the two frames to find the volume flux.  The same method was continued for the entire  sequence of contour pairs to obtain the volume fluxes as a function of time.

\section{Cavity collapse}\label{bubble_collapse}
\begin{figure}
    \centering
    \begin{subfigure}{0.19\textwidth}
        \includegraphics[width=\textwidth]{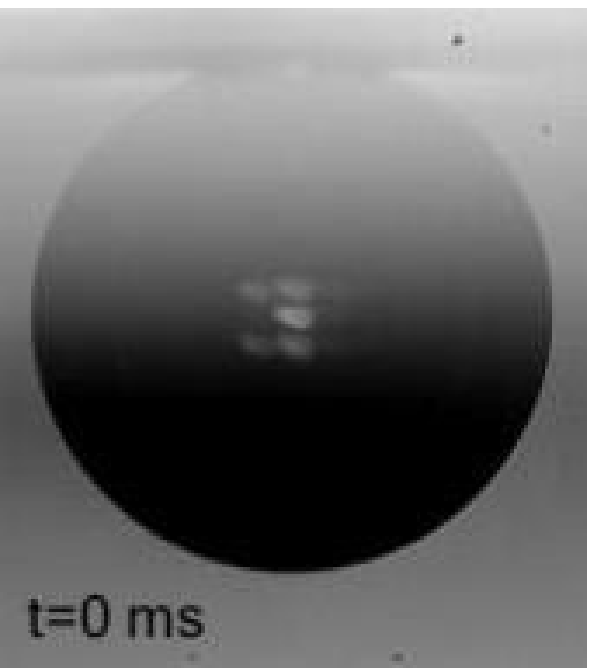}
        \caption{}
        \label{}
    \end{subfigure}
    \begin{subfigure}{0.19\textwidth}
       \includegraphics[width=\textwidth]{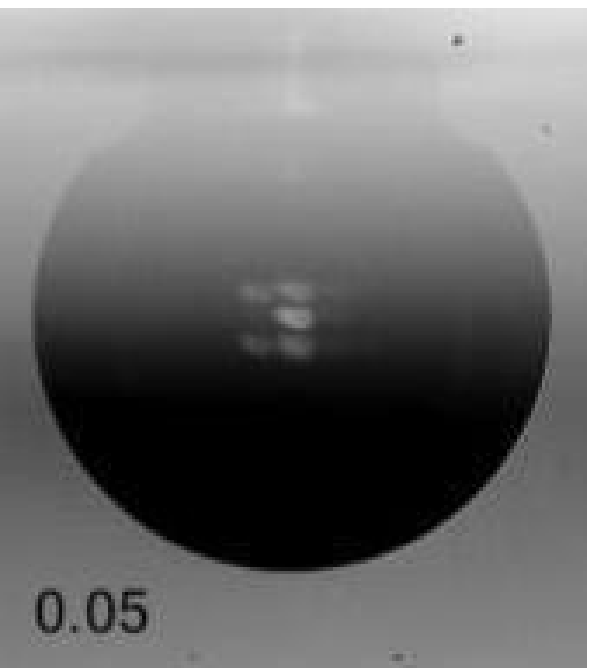}
        \caption{}
        \label{}
    \end{subfigure}
     \begin{subfigure}{0.19\textwidth}
       \includegraphics[width=\textwidth]{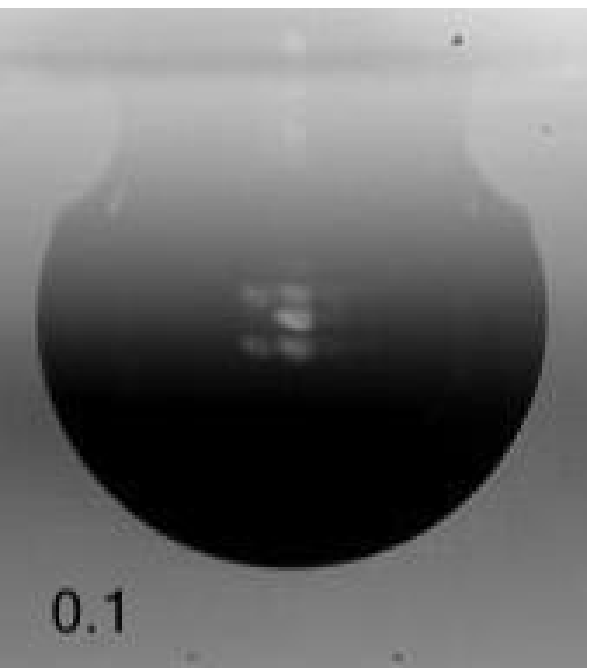}
        \caption{}
        \label{}
    \end{subfigure}
     \begin{subfigure}{0.19\textwidth}
       \includegraphics[width=\textwidth]{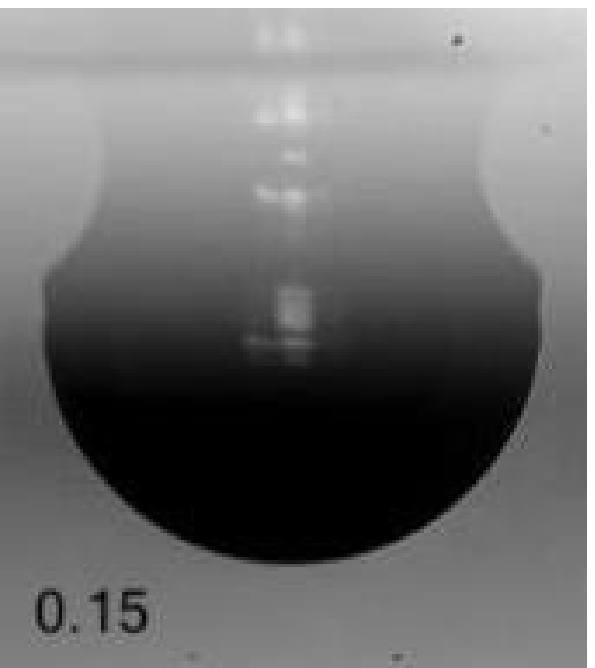}
        \caption{}
        \label{}
    \end{subfigure}
    \begin{subfigure}{0.19\textwidth}
       \includegraphics[width=\textwidth]{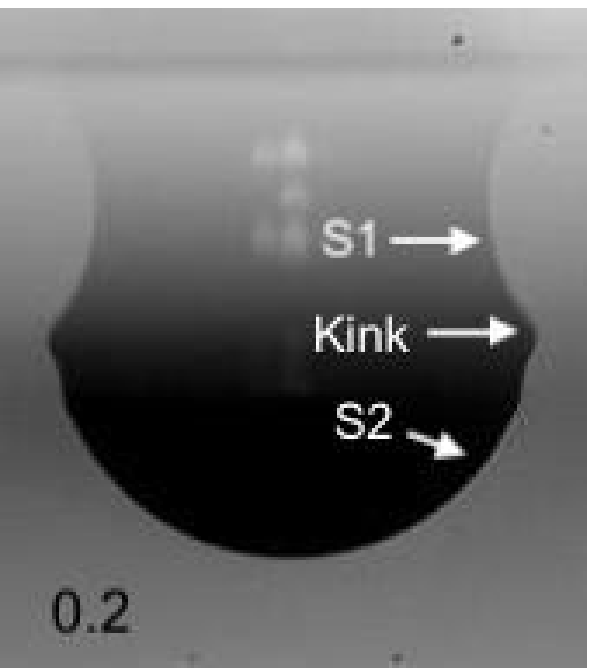}
        \caption{}
        \label{img188}
    \end{subfigure}
    \begin{subfigure}{0.19\textwidth}
       \includegraphics[width=\textwidth]{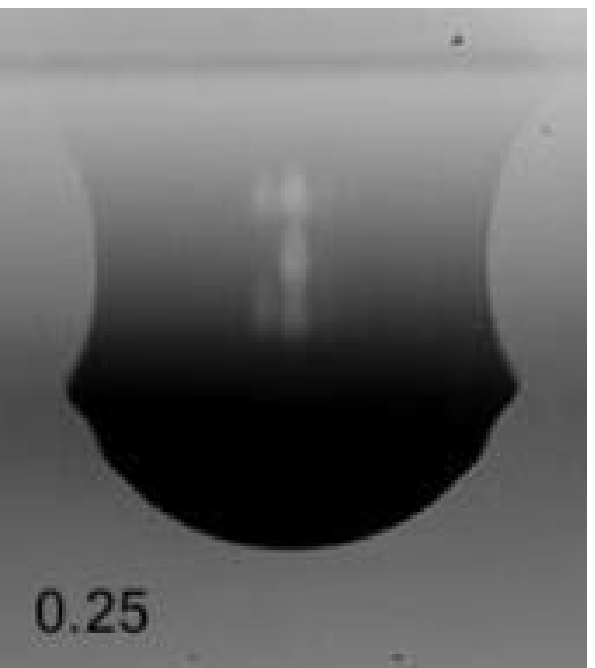}
        \caption{}
        \label{}
    \end{subfigure}
    \begin{subfigure}{0.19\textwidth}
       \includegraphics[width=\textwidth]{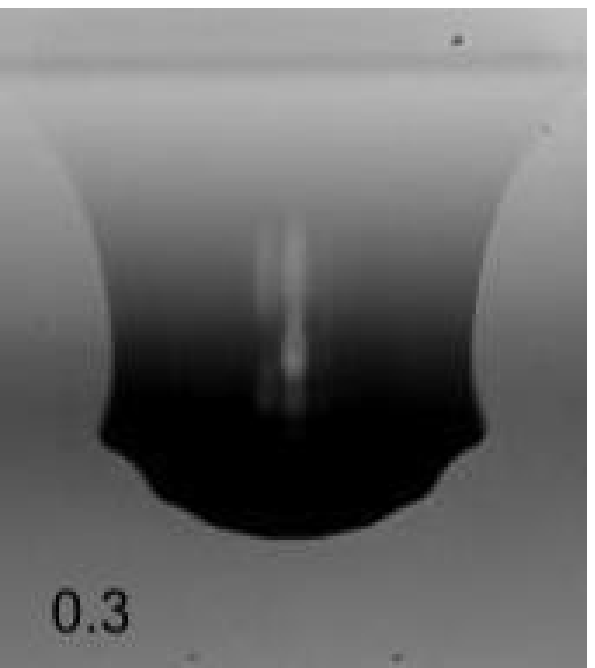}
        \caption{}
        \label{}
    \end{subfigure}
    \begin{subfigure}{0.19\textwidth}
       \includegraphics[width=\textwidth]{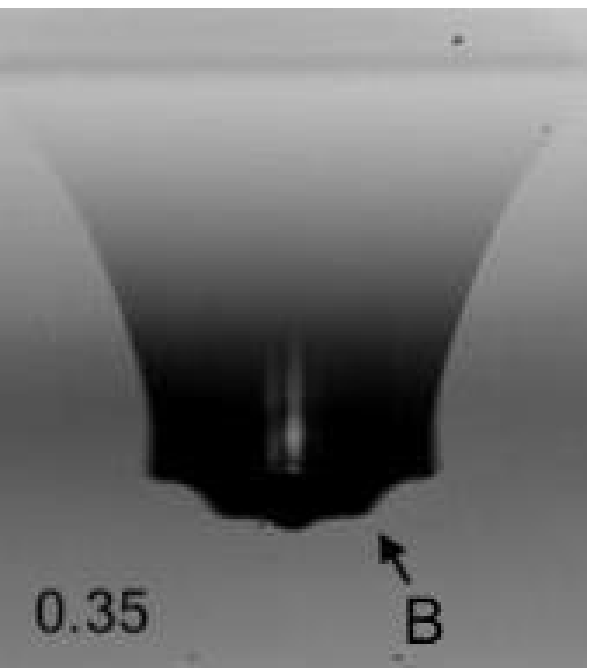}
        \caption{}
        \label{img77nw}
    \end{subfigure}
    \begin{subfigure}{0.19\textwidth}
       \includegraphics[width=\textwidth]{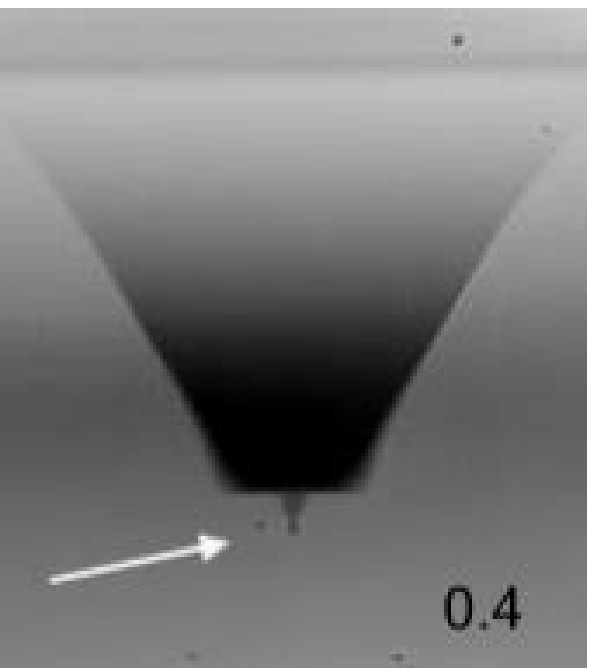}
        \caption{}
        \label{}
    \end{subfigure}
    \begin{subfigure}{0.19\textwidth}
       \includegraphics[width=\textwidth]{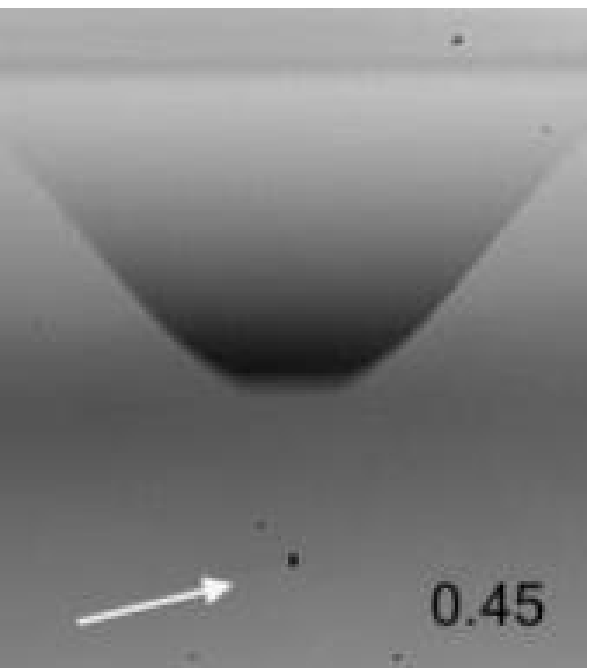}
        \caption{}
        \label{}
    \end{subfigure}
     \caption{Image sequence showing the stages of cavity collapse in a low viscosity fluid, showing the presence of precursory capillary waves. The  bubble is of radius  $R=0.47$mm in water ($Bo=0.03$) ($Oh=0.0055$). Bubble pinch-off from wave focusing,  creating a downward gas jet of radius  $8.6 \mu$m,  can also be seen in (h) to (i).  The width of each image is 0.97 mm. Movie~1.}
    \label{satellitebubble}
\end{figure}

\begin{figure}
    \centering
		\begin{subfigure}{0.19\textwidth}
        \includegraphics[width=\textwidth]{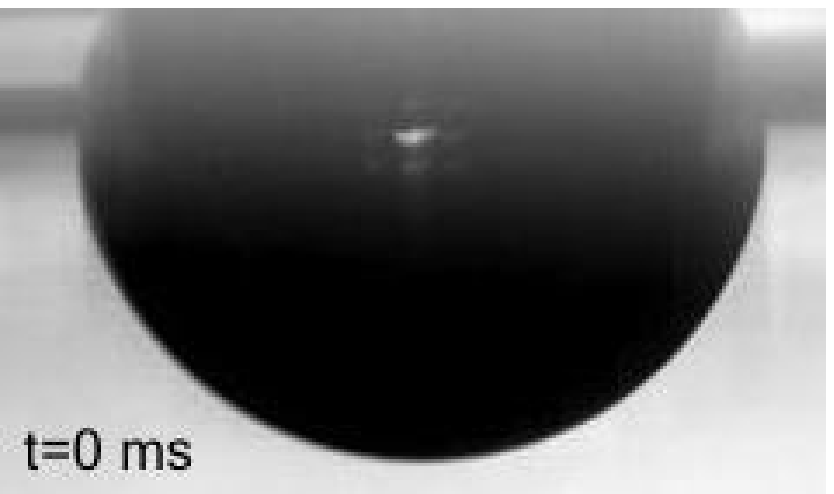}
        \caption{}
        \label{}
    \end{subfigure}
    \begin{subfigure}{0.19\textwidth}
        \includegraphics[width=\textwidth]{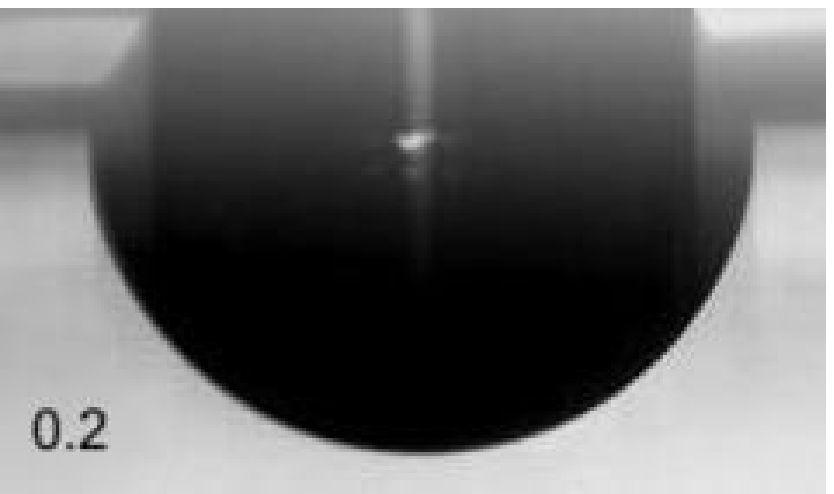}
        \caption{}
        \label{}
    \end{subfigure}
    \begin{subfigure}{0.19\textwidth}
       \includegraphics[width=\textwidth]{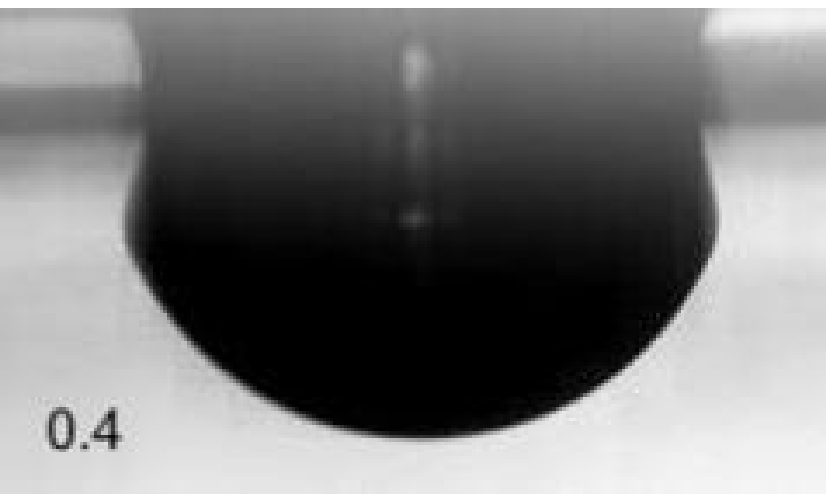}
        \caption{}
        \label{}
    \end{subfigure}
     \begin{subfigure}{0.19\textwidth}
       \includegraphics[width=\textwidth]{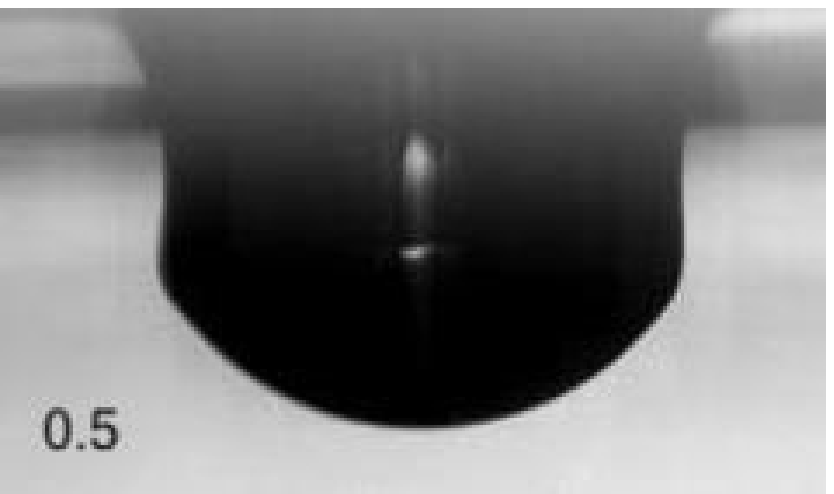}
        \caption{}
        \label{}
    \end{subfigure}
     \begin{subfigure}{0.19\textwidth}
       \includegraphics[width=\textwidth]{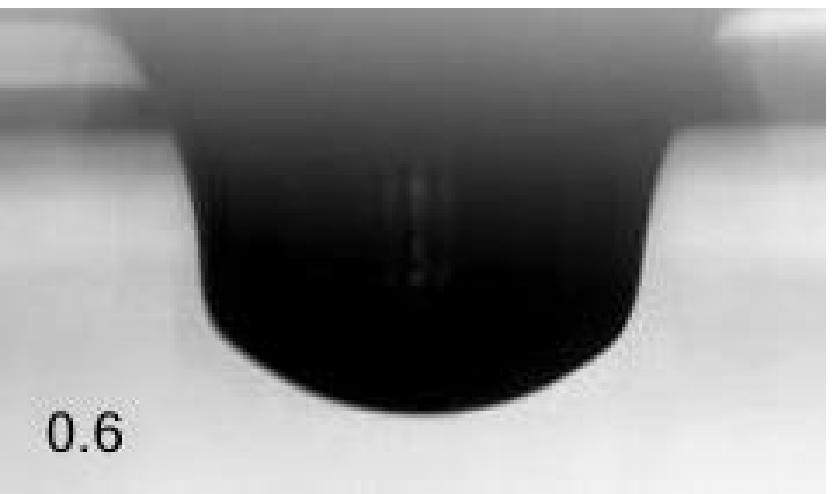}
        \caption{}
        \label{}
    \end{subfigure}
    \begin{subfigure}{0.19\textwidth}
       \includegraphics[width=\textwidth]{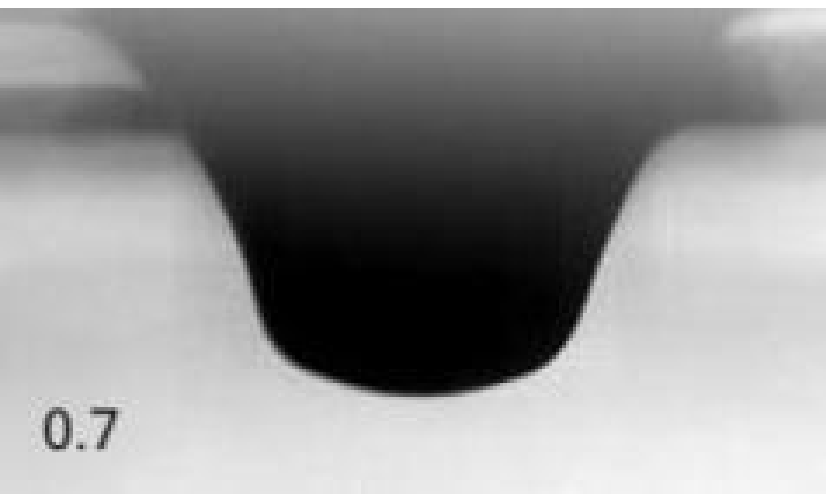}
        \caption{}
        \label{}
    \end{subfigure}
    \begin{subfigure}{0.19\textwidth}
       \includegraphics[width=\textwidth]{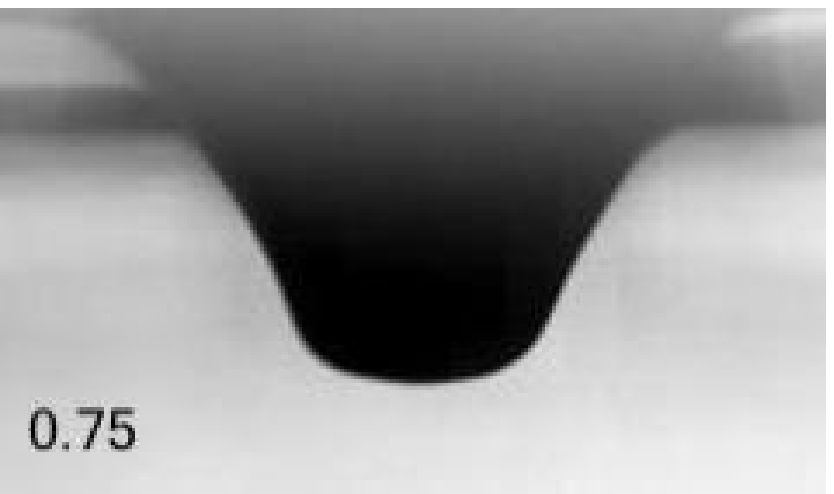}
        \caption{}
        \label{}
    \end{subfigure}
    \begin{subfigure}{0.19\textwidth}
       \includegraphics[width=\textwidth]{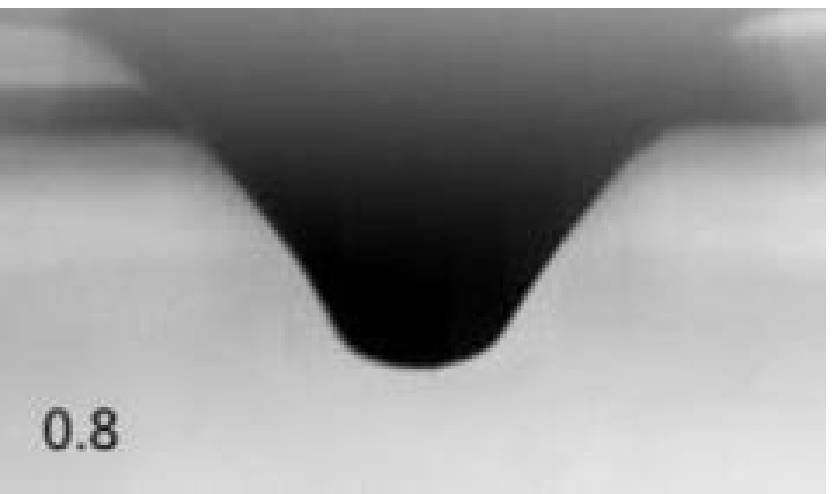}
        \caption{}
        \label{}
    \end{subfigure}
    \begin{subfigure}{0.19\textwidth}
       \includegraphics[width=\textwidth]{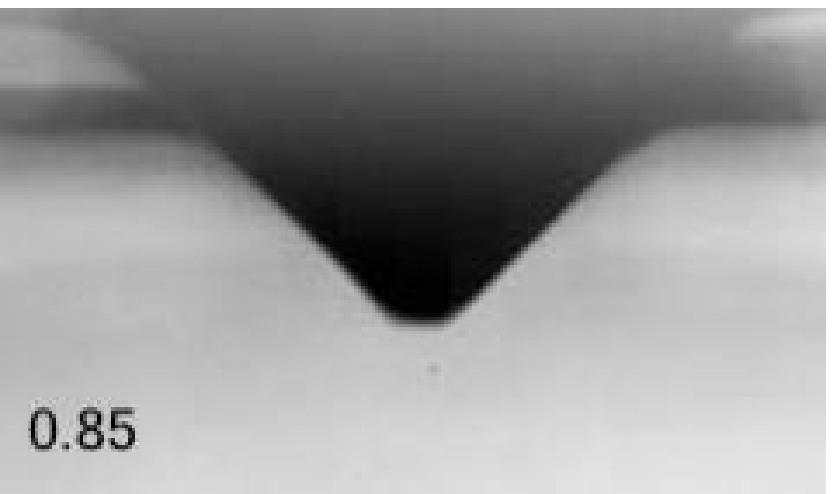}
        \caption{}
        \label{}
    \end{subfigure}
     \caption{Image sequence showing the stages of cavity collapse due to a $R=0.7$mm bubble ($Bo=0.08$) in a high-viscous fluid (GW55, $Oh=0.034$) that is free of the  precursory capillary waves. The image width is $1.7$mm. Movie~2.}
    \label{cavitycollapsegw55_CR}
\end{figure}

 Figures~\ref{satellitebubble} shows a sequence of the stages of a bubble collapse at the free-surface for a low viscosity fluid (water, $Oh=0.0055$).
 The corresponding stages for a high viscosity fluid (GW55, $Oh=0.034$) are shown in  figure~\ref{cavitycollapsegw55_CR}.
 In both cases,  an axisymmetric  kink (figure~\ref{img188})  is seen travelling from the  cavity top to the bottom-most part of the cavity, where the liquid converges.  In addition to a much faster collapse in the low viscosity case (figure~\ref{satellitebubble}),  precursor capillary waves are seen moving ahead of the kink   (see B in figure~\ref{img77nw}). We observe these capillary waves only when  $Oh<0.02$, as it is the case in figure~\ref{satellitebubble},  resulting in a sharp front edge of the kink, as can be seen in figure~\ref{satellitebubble}. When $Oh>0.02$, as shown in   figure~\ref{cavitycollapsegw55_CR}, these  precursory capillary waves are fully damped by the viscous effects, resulting in a more rounded front edge of the kink. As can be seen from figures~\ref{satellitebubble} and \ref{cavitycollapsegw55_CR}, the edge of the kink travels along the cavity surface, while the cavity itself is shrinking normal to its surface. Thus, at any instant, the kink has velocities tangential and normal to the cavity surface,  up to the flow focusing at the cavity bottom. We analyse these velocities  in detail in the following sections.

 \subsection{{Tangential velocity of the kink} }\label{Uttext}
\begin{figure}
 \centering
  \includegraphics[width=0.95\textwidth]{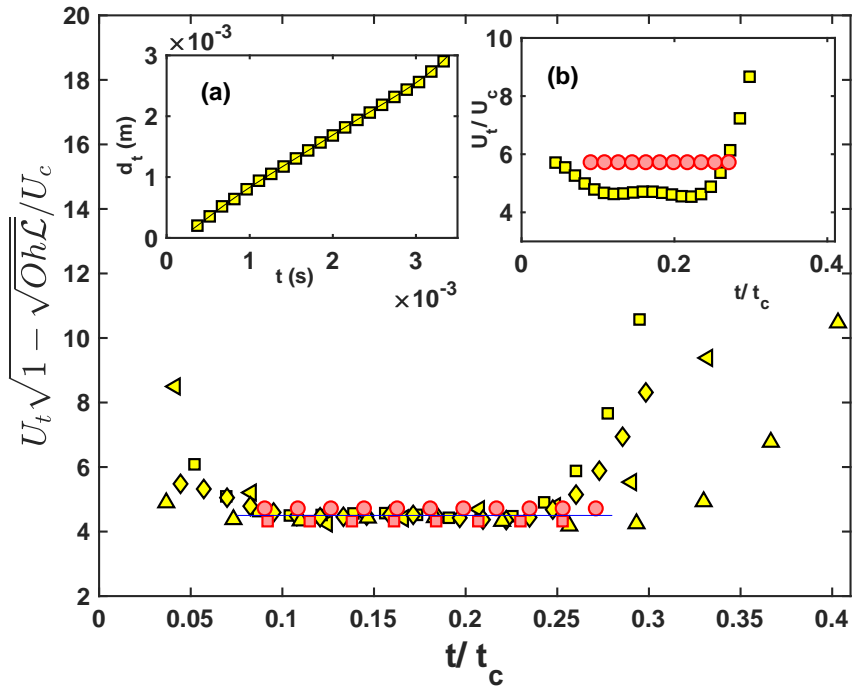}
  \caption{Normalised tangential velocity of the kink, accounting for the wave resistance of precursory capillary waves, ${\mathcal{W}}_R(Oh,Bo)$,  and the path correction due to gravity effects, $\mathscr{L}(Bo)$, as in (\ref{wave_resi_4}), plotted as a function of the dimensionless time $t/t_c$, {for bubbles of $0.001 <Bo < 1$ and $0.001<Oh<0.05$}. ${\color{yellow}\blacktriangle}$, $R=0.175$mm   ($Bo=4.2 \times 10^{-3}$, $Oh=0.0099$); ${\color{yellow}\blacktriangleleft}$, $R=0.47$mm  ($Bo=3 \times 10^{-2}$, $Oh=0.0055$); ${\color{yellow}\blacksquare}$, $R=1.74$mm  ($Bo=4.1 \times 10^{-1}$, $Oh=0.0028$) and ${\color{yellow}\blacklozenge}$ $R=2.15$mm  ($Bo= 6.3 \times 10^{-1}$, $Oh=0.00255$). Aforementioned  data are from water.  Data with GW72 are: ${\color{red}\blacksquare}$ $R=1.59$mm  ($Bo= 4.8 \times 10^{-1}$, $Oh= 0.0481$); ${\color{red}\bullet}$ $R=2.02$mm  ($Bo= 7.7 \times 10^{-1}$, $Oh=0.0427$).  ---, $U_t\sqrt{1-\sqrt{Oh\mathscr{L}}}/U_c~=4.5$. In inset (a), the cumulative distance, $d_t$, travelled by the kink in the tangential direction is plotted verses time for $R=2.14$mm in water with, ---, the polynomial fit used for calculating $U_t$. The inset (b) shows the data offset between water and GW72 when precursory capillary wave effects are not taken into account.}
    \label{Vt_inst1}
\end{figure}

The tangential velocities  $U_t$ of the kink  were measured as discussed in \S~\ref{expsetup}. Inset (b) in figure~\ref{Vt_inst1} shows the variation of the dimensionless  tangential velocity ($U_t/U_c$)  with the dimensionless time ($t/t_c$),  for  bubbles of similar $Bo$ in GW72, $Oh=0.0427$ (red circle), and in water, $Oh=0.0028$ (yellow square). 
$U_t$ is observed to be constant, except at the beginning and the end of the collapse, and scale with the capillary velocity  $U_c=\sqrt{\sigma/\rho R}$, in a way similar to the observations of  \cite{Sangeeth2015207,krishnan_hopfinger_puthenveettil_2017} and \cite{gordillo2019capillary}.  However, for $Oh=0.0028$,  where  precursory capillary waves occur ahead of the kink, as shown in figure~\ref{satellitebubble}, the values of $U_t/U_c$ are around $40\%$ lower compared with those at $Oh=0.0427$, where the  precursor capillary waves are fully damped (figure~\ref{cavitycollapsegw55_CR}). We observe this behaviour with all the  bubbles when $Oh<0.02$. Similar diminishing velocity of the kink in the presence of precursory capillary waves is clearly seen in the velocity data of \cite{ji2021compound}  \citep[see][figure~5]{ji2021compound}  for the bursting of bubbles in  oil covered water surface, where the oil layer covering the kink  enhances the damping of precursory capillary waves. Thus the capillary velocity scale alone does not collapse the tangential velocity data for different viscosity fluids, possibly due to the effect of precursory capillary waves on $U_t$.  A new scaling relation for $U_t$ is therefore needed to account for the effect of viscous damping of  the  precursor capillary waves, and a  possible (weak) gravity effect. Using an energy balance at the kink, we now obtain such a scaling relation that collapses the tangential velocity data.

\subsubsection{Energy balance {at the kink} }\label{wave_res_text}
The retraction of the rim right after the film rupture provides the kinetic energy associated with the kink movement. Since the kink moves with constant velocity $U_t$, as seen in  \S~\ref{Uttext} and figure~\ref{Vt_inst1}, we assume a steady state balance of the energy of the kink movement. Assuming  $R$  to be the characteristic length  in the azimuthal and the  vertical directions of the collapsing cavity, and the amplitude of $a$  to be of the order of the wavelength of the precursory capillary waves, $a\simeq \lambda$, by balancing the kinetic energy of the kink, having a mass $\rho aR^2$, with the energy spent on generating the precursory capillary waves,
\begin{equation}
    \frac{1}{2}\rho a R^2  {U_t}^2  =   \alpha_1 \frac{1}{2}\rho a R  \lambda {U_t}^2 + \alpha_2 \sigma a R , 
    \label{wave_resi_2}
\end{equation}
  where $\alpha_1$, $\alpha_2$ are constant prefactors.  Rearranging  (\ref{wave_resi_2}), we obtain the Weber number of cavity collapse  in the form,
\begin{equation}
  We_c = \frac{\rho {U_t}^2 R}{\sigma} = \left(\frac{U_t}{U_c}\right)^2 = \frac{2 \alpha_2}{1 - \alpha_1 {\lambda}/{R}}.
    \label{wave_resi_3}
\end{equation}
  The expression (\ref{wave_resi_3}) quantifies the reduction in $U_t/U_c$, shown in the inset (b) of figure~\ref{Vt_inst1},  due to the presence of precursory capillary waves. The dimensionless wavelength $\lambda/R$ of the dominant precursory capillary wave in (\ref{wave_resi_3}) depends on the total time of cavity collapse, which, as we show later in \S~\ref{wavlength_tbc_Ut_final_textpart}, depends on $Oh$ and $Bo$. Then, (\ref{wave_resi_3}) can be written as  
  \begin{equation}
      U_t=\alpha_3~ U_c ~{\mathcal{W}}_R(Oh,Bo),
      \label{UtUcWR_1}
  \end{equation}
  where $\alpha_3=\sqrt{2 \alpha_2}$ and
  \begin{equation}
    {\mathcal{W}}_R(Oh,Bo)=1/\sqrt{1-\alpha_1\lambda/R}  
    \label{WR_defn}
  \end{equation}
  is the wave resistance factor that accounts for the reduction in $U_t$ due to the precursory capillary waves. ${\mathcal{W}}_R(Oh,Bo)$ depends on $\lambda/R$, which in turn depends on the total time of cavity collapse, $t_{bc}$, since viscous damping during $t_{bc}$ affects $\lambda/R$. We now discuss the dependency of $t_{bc}$ on $Oh$ and $Bo$, which allows us to get the dependency of $\lambda/R$ on $Oh$ and $Bo$ and thereby an expression for ${\mathcal{W}}_R(Oh,Bo)$.
  
  \subsubsection{Total time of  cavity collapse $t_{bc}$}\label{tbcsection}
\begin{figure}
 \centering
  \includegraphics[width=0.95\textwidth]{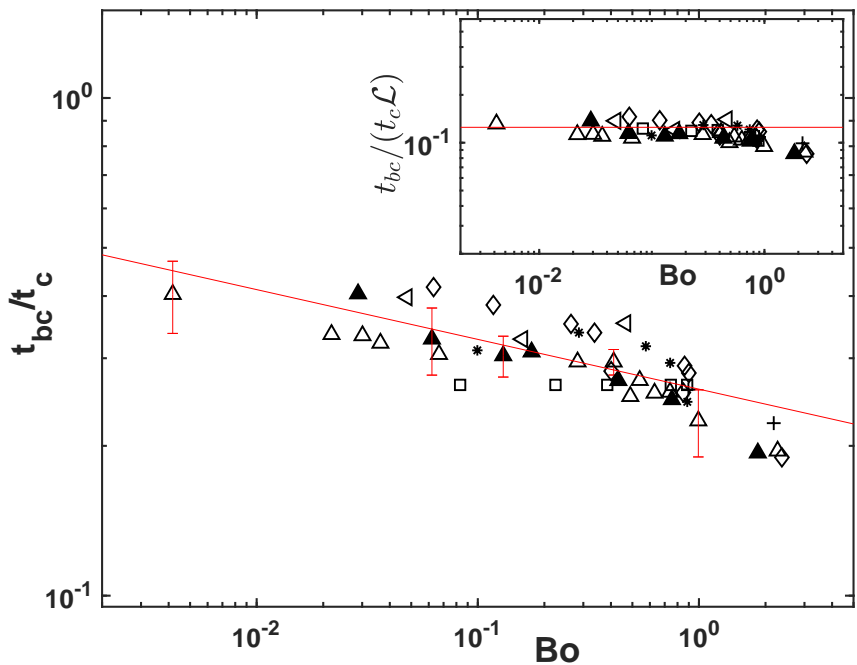}
  \caption{The  effect of Bond number on the dimensionless total time of cavity collapse  $t_{bc}/t_c$, where $t_c$ is the capillary time scale.  ---, ${t_{bc}/t_{c}=0.26~ Bo^{-0.1}}$. In the inset, dimensionless total time of cavity collapse, accounting for the path correction due to gravity,  ${t_{bc}}/(t_c~\mathscr{L})$, is plotted as a function of $Bo$.  ---, ${t_{bc}}/(t_c~\mathscr{L})=~0.13$ (\ref{tbcRreqn}). $\triangle$, Water; $\triangleleft$, ethanol; $\blacktriangle$, GW48 (30$^{\circ}$C); $*$, GW68; $\lozenge$, GW72; $+$, 2-propanol;  $\square$, GW55.}
    \label{tbc_Bo_effect}
\end{figure}

 Since the time taken for the disintegration of the film is negligible \citep{duchemin2002jet}, we consider the time at which the retracting rim has reached the outer edge of the film, at $R_r$, to be the reference time $t=0$ (see figure~\ref{DeDnb_labelled}). The time from $t=0$  to the stage where the cavity has become conical, just before the ejection of the jet (figure~\ref{satellitebubble}i), is measured as the total time of cavity collapse, $t_{bc}$. In figure~\ref{tbc_Bo_effect},  $t_{bc}$,  normalised by the capillary time scale $t_{c}=\sqrt{\rho R^{3}/\sigma}$, is plotted  as a function of $Bo$. 
 The experimental data indicate a gravity dependency of $t_{bc}/t_c$ in the form,  
 \begin{equation}
     \frac{t_{bc}}{t_{c}}=0.26~ Bo^{-0.1}.
     \label{tbcBo0p1eqn}
 \end{equation}
 
This Bond number dependence of $t_{bc}$ can be physically explained by evaluating the time taken by the kink to travel along the cavity boundary. The length of the  path travelled by the kink ($s_{bc}$) along the cavity surface, from the rim till the bottom of the cavity (see figure~\ref{DeDnb_labelled}),  is  a function of $Bo$  because  the static shape of the free-surface bubble depends on $Bo$. For $Bo\leq 1$, 
\begin{equation}
     s_{bc}=(\pi R - s_f)Oh^n \approx  (\pi R - R_r)Oh^n,
     \label{sbcBo0p1eqn}
 \end{equation}
 where $s_f=R \theta$ (see figure~\ref{DeDnb_labelled}), with $\theta=R_r/R$ for small $\theta$ and $R_r$ is the rim radius (see figure~\ref{B6}), which is a function of $Bo$ \citep{puthenveettil2018shape}.  The factor $Oh^n$, where $n$ is a positive exponent, appears in (\ref{sbcBo0p1eqn}) because the bottom radius of the conical cavity $r_b$ (see  figure~\ref{DeDnb_labelled}) depends on $Oh$. The occurrence of capillary waves, which increase $r_b$ \citep{gordillo2019capillary}, depends on $Oh$.  
 After substituting $s_{bc}$ from (\ref{sbcBo0p1eqn}) in $t_{bc}\approx s_{bc}/U_t$,  with   $U_t$ given by (\ref{UtUcWR_1}), we obtain,
\begin{equation}
\label{tbcRreqn_full}
{t_{bc}}~\approx~\frac{1}{\alpha_3}~t_c \mathscr{L} \zeta,
\end{equation}
where $\zeta=Oh^n/{\mathcal{W}}_R$, with ${\mathcal{W}}_R$ given by (\ref{WR_defn}) and 
\begin{equation}
\label{LBo_def}
    \mathscr{L}(Bo)=\pi -{R_r}/{R}
\end{equation}
 is the path correction term  that accounts for the gravity dependence of the path length $s_{bc}$  travelled by the kink.  In (\ref{LBo_def}), the dimensionless rim radius  
 \begin{equation}
    R_r/R=\sqrt{4/3-2(1/Bo+1/Bo^2)+\sqrt{-4/3Bo^2+8/Bo^3+4/Bo^4}},
    \label{Rreqn_paper}
 \end{equation}
 when $Bo<1$ \citep{puthenveettil2018shape}.

 Equation (\ref{tbcRreqn_full}) delineates the capillary effects on the total time of cavity collapse through $t_c$ while the gravity and viscous effects through  $\mathscr{L}(Bo)$ and  $\zeta$, respectively.  The inset in  figure~\ref{tbc_Bo_effect} shows that the measured values of $t_{bc}/(t_c~\mathscr{L})$ collapse onto 
\begin{equation}
\label{tbcRreqn}
{t_{bc}}~\approx~0.13~t_c~\mathscr{L}, 
\end{equation}
for $0.001<Bo<1$ and fluids of various viscosity. The deviation of the data from (\ref{tbcRreqn}), when $Bo>1$, occurs because $s_{bc}$ starts to deviate from (\ref{sbcBo0p1eqn}) and (\ref{Rreqn_paper}) due to increasing deviations of the shape of the cavity from that of a truncated sphere. Equations (\ref{tbcRreqn_full}) and (\ref{tbcRreqn}) imply that $Oh^n/(\alpha_3{\mathcal{W}}_R)=0.13$ for the present range of $0.001<Oh<0.1$. Then, the increase in $s_{bc}$ at larger $Oh$ due to decreasing $r_b$ (see (\ref{sbcBo0p1eqn})) seems to be offset by increasing velocities due to increased damping of precursory capillary waves (see (\ref{wave_resi_3})), so that $t_{bc}$ becomes independent of $Oh$, as given by (\ref{tbcRreqn}). The total time of cavity collapse then follows a capillary time scale $t_c$, modified by the term $\mathscr{L}$, which depends on $Bo$ through (\ref{LBo_def}) and (\ref{Rreqn_paper}), with negligible dependence on viscosity.  

\subsubsection{Wavelength of precursor capillary waves and scaling of $U_t$}\label{wavlength_tbc_Ut_final_textpart}
The $Bo$ dependence of $t_{bc}$ given by  (\ref{tbcRreqn}) requires modification of the wave damping scaling relation $\lambda/R \propto \sqrt{Oh}$ presented in \cite{krishnan_hopfinger_puthenveettil_2017}, which was based on   $t_{bc}\approx 0.3 t_c$, proposed by \cite{Sangeeth2015207}. It has been shown that the amplitudes of the capillary waves fall off exponentially in the form $a/a_0=e^{-\kappa~t}$, where $a_0$ and $a$ are, respectively, an initial and later wave amplitude, with $\kappa~=~8\pi^2 \mu/\rho \lambda^2$ being the wave damping coefficient \citep{lighthill2001waves}. The waves can be considered fully damped at the end of the cavity collapse time $t_{bc}$, when 
\begin{equation}
\kappa t_{bc} = \frac{8\pi^2 \mu t_{bc}}{\rho \lambda^2}=4.
\label{wavedampingcondn}
\end{equation}
Substituting (\ref{tbcRreqn}) for $t_{bc}$ in (\ref{wavedampingcondn}), and rearranging, we obtain {the dimensionless wave length that is damped in the time $t_{bc}$ as,} 
\begin{equation}
\frac{\lambda}{R}= {c_2}\sqrt{Oh{\mathscr{L}}},
\label{lambda_Oh_Bo_new}
\end{equation}
where $c_2=0.5\pi$.  The experimental values corresponding to $Bo\approx 0.37$ are typically $\lambda/R\approx0.4$, which,  according  to (\ref{lambda_Oh_Bo_new}),  when ${\mathscr{L}}\approx2.5$, requires $Oh\approx 0.02$, the value below which we observe precursory capillary waves.   The gravity dependency of the path correction ${\mathscr{L}}$ is given by (\ref{LBo_def}) and (\ref{Rreqn_paper}). Since $R_r/R$ increases with increasing $Bo$, as given by (\ref{Rreqn_paper}),  ${\mathscr{L}}$ decreases when $Bo$ is increased. Thus, when  $Bo$ is large, $Oh$ needs to be larger for the waves to be damped in the time $t_{bc}$. 

Substituting $\lambda/R$ from (\ref{lambda_Oh_Bo_new}) in (\ref{UtUcWR_1}) and rearranging, we get  
\begin{equation}
   U_t~\approx~\frac{\alpha_3~U_c}{\sqrt{1 - c_2 \alpha_1 \sqrt{Oh {\mathscr{L}}}}}.
    \label{wave_resi_4}
\end{equation}

In figure~\ref{Vt_inst1}, the dimensionless tangential velocity  $U_t\sqrt{1 - c_2 \alpha_1 \sqrt{Oh {\mathscr{L}}}}/U_c$ is plotted against the dimensionless time $t/t_c$ for bubbles in water (yellow symbols) and GW72 (red symbols) in the range $2 \times 10^{-3} < Bo < 1$ and $0.001<Oh<0.05$.  The data collapse onto 
\begin{equation}
\label{tangU1}
 \frac{U_t\sqrt{1 - \sqrt{Oh {\mathscr{L}}}}}{U_c} = 4.5,
  \label{wave_resi_datamatch}
\end{equation} 
for $0.05 < t/t_c < 0.3$,  the uniform phase of tangential  motion,  implying that ${\alpha_1=1/c_2}$ and ${\alpha_3=4.5}$. 
The relation (\ref{wave_resi_datamatch}), in its zero $Oh$ limit, matches with the relation for the absolute velocity (see figure~\ref{cavitycollapse_labelled}) of the dominant capillary wave, $U_{abs}\simeq{5U_c}$, proposed by \cite{gordillo2019capillary},  based on their  numerical simulation in the vanishing Bond number limit $Bo\ll0.1$.   At finite $Bo$ and $Oh$, (\ref{wave_resi_datamatch}) captures the complex dependence of the kink velocity on $Oh$ and $Bo$ that occurs  through the damping of the precursory capillary waves. 

Comparing (\ref{wave_resi_datamatch}) with (\ref{UtUcWR_1}), shows that the wave resistance factor in (\ref{UtUcWR_1}) is of the form, 
\begin{equation}
\label{WREqn}
 {\mathcal{W}}_R(Oh,Bo)= \frac{1}{\sqrt{1-~ \sqrt{Oh {\mathscr{L}}}}}.
\end{equation} 
Then, the final scaling of the tangential velocity of the kink is ${U_{t}}~\approx~4.5~U_c~{\mathcal{W}}_R$, where ${\mathcal{W}}_R$ is given by (\ref{WREqn}).

The constancy of $U_t$ with respect to time, seen in figure~\ref{Vt_inst1}, could also be understood in terms of the phase velocity of the precursory capillary waves. The kink produces a wave disturbance, of wave length $\lambda$, at the cavity surface, which propagates like a capillary wave with a phase velocity $c_p=(2\pi)^{1/2}\sqrt{{\sigma}/{\rho \lambda}}$. Substituting $\lambda$ from (\ref{lambda_Oh_Bo_new}) in this relation gives
\begin{equation}
\label{cp_eqn}
    c_p =2.8U_c (Oh{\mathscr{L}})^{-1/4},
\end{equation}
 which is close to (\ref{wave_resi_datamatch}),  although the dependency on viscosity and gravity shown by (\ref{wave_resi_datamatch}) is not fully captured by $c_p$. However, the important point is that the phase velocity of the precursory capillary wave does give an argument for $U_t$ being constant in time, as seen in figure~\ref{Vt_inst1}. 

 \subsection{Shrinking of the cavity boundary in the normal direction}\label{Normalvel_shrinking}
\begin{figure}
    \centering
         \begin{subfigure}{0.29\textwidth}
        \includegraphics[width=\textwidth]{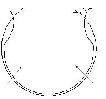}
        \caption{}
        \label{normalmotion_superimposed}
    \end{subfigure}
    \begin{subfigure}{0.30\textwidth}
        \includegraphics[width=\textwidth]{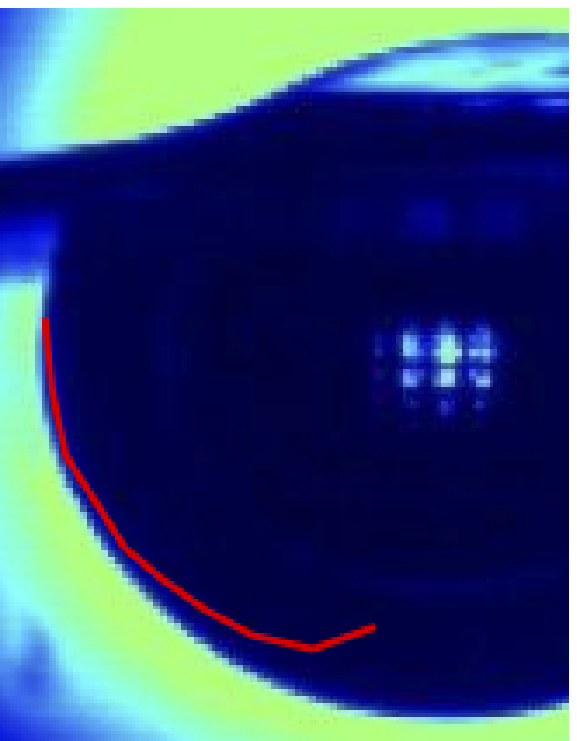}
        \caption{}
        \label{water_le_trj}
    \end{subfigure}
		\begin{subfigure}{0.285\textwidth}
        \includegraphics[width=\textwidth]{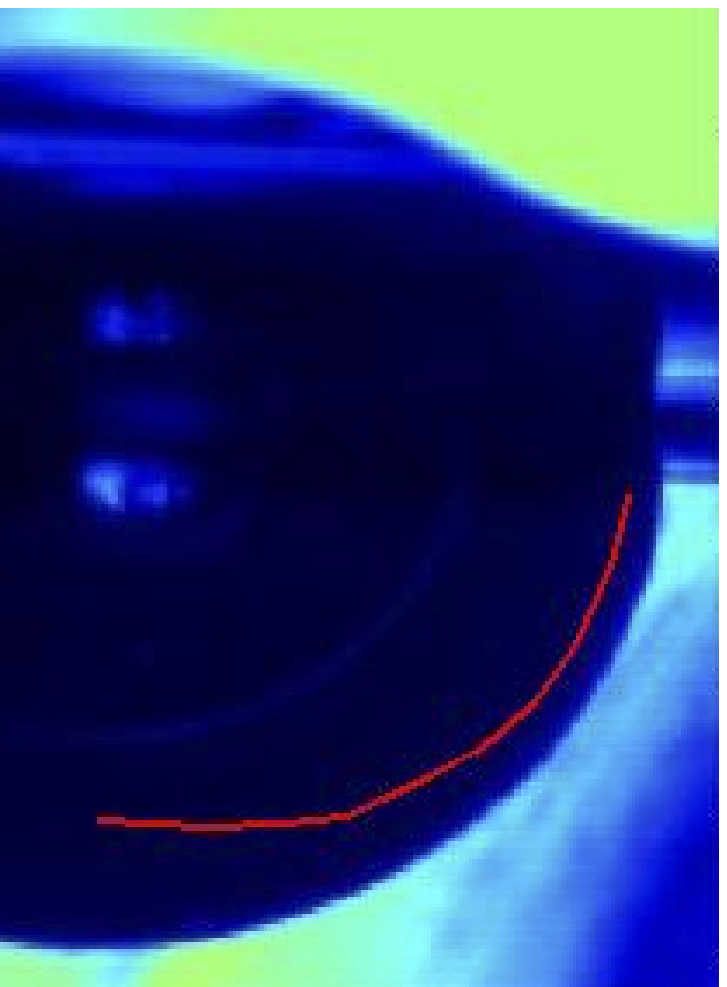}
        \caption{}
        \label{gw72_le_trj}
    \end{subfigure}
     \caption{Shrinking of the cavity in the normal direction.  (a),  contours of the cavity at two time instances, superimposed for a bubble of $R=0.5$mm in water. Arrows show the  shrinking of the cavity boundary below the kink rim. The time gap between the two contours is  $0.15$ms. The (red) lines superimposed over the static shapes of the bubbles in (b) and (c) indicate the trajectories of the kinks extracted from their cavity collapse. The bubbles in (b) and (c) are of similar size but differ significantly in $Oh$; (b), $R=2.15$mm, $Bo=0.63$, $Oh=0.00255$ in water; (c), $R=2$mm, $Bo=0.77$, and  $Oh=0.0427$ in GW72. The width of the images are $5.43$mm and $2.85$mm.}
    \label{tangvelandtrajectories}
\end{figure}

In figure~\ref{normalmotion_superimposed} the initial cavity boundary (the bubble boundary) at $t=0$ is compared with that at a later instant ($t=0.15$ms) for a bubble of $R=0.5$mm in water.
The cavity contours  clearly indicate the retraction of the cavity in the normal direction everywhere below the kink rim, as indicated by the arrows. This cavity shrinkage is due to the sudden reduction of the gas pressure in the cavity after rupture of the surface film, leading to an imbalance with the surface tension force, which scales as  ${\sigma/R}$.

In figures \ref{water_le_trj} and \ref{gw72_le_trj}, the trajectories of the kink are indicated by the continuous (red) lines in the images of a bubble of $R=2.15$mm in water and of a bubble of $R=2$mm in GW72, respectively. It is seen that the extent of shrinkage, \emph{i.e.,} the gap between the initial cavity contour and the red line, is larger for the bubble in the viscous fluid GW72 than it is in water.   It is also seen that in the water bubble in figure~\ref{water_le_trj} that the kink  undergoes a sudden jump towards the end, while the trajectory of the kink in GW72 (figure~\ref{gw72_le_trj}) is smooth throughout the collapse. This sudden rise in velocity of the kink is a feature observed for bubbles in low viscosity fluids  of $Oh<0.02$, where precursory capillary waves are present. 
\begin{figure}
 \centering
  \includegraphics[width=0.55\textwidth]{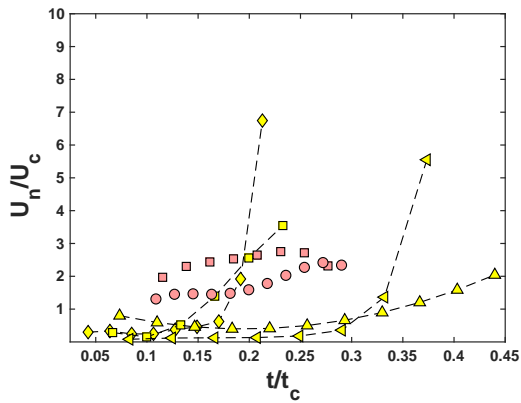}
	\caption{The dimensionless normal velocity of the leading edge  $U_n/U_c$ as a function of the dimensionless time $t/t_c$. The symbols are the same as that in figure~\ref{Vt_inst1} namely:  ${\color{yellow}\blacktriangle}$, $R=0.175$mm   ($Bo=4.2 \times 10^{-3}$, $Oh=0.0099$); ${\color{yellow}\blacktriangleleft}$, $R=0.47$mm  ($Bo=3 \times 10^{-2}$, $Oh=0.0055$); ${\color{yellow}\blacksquare}$, $R=1.74$mm  ($Bo=4.1 \times 10^{-1}$, $Oh=0.0028$) and ${\color{yellow}\blacklozenge}$ $R=2.14$mm  ($Bo= 6.3 \times 10^{-1}$, $Oh=0.00255$). Aforementioned  data are from water.  Data with GW72 are: ${\color{red}\blacksquare}$ $R=1.59$mm  ($Bo= 4.8 \times 10^{-1}$, $Oh= 0.0481$); ${\color{red}\bullet}$ $R=2.02$mm  ($Bo= 7.7 \times 10^{-1}$, $Oh=0.0427$).}
  \label{Vn_curvefit}
\end{figure}

In figure~\ref{Vn_curvefit}, the  velocity  of the kink in the  direction normal to the cavity boundary, $U_n$, normalised with the capillary velocity, $U_c$, is plotted  as a function of  the dimensionless time $t/t_c$.  The scaled velocities in the viscous fluid (GW72, ${Oh>0.02}$),  are higher than in water (${Oh<0.02}$);  clearly, a capillary velocity scaling alone, as in the figure, does not collapse the normal velocities. The velocity data also show an increase with time, indicating a weak acceleration, except in water, where toward flow-focusing, $U_n/U_c$ values  increase due to the presence of precursory waves, discussed in \S~\ref{Uttext}. 

It needs to be noted that a ${\mathcal{W}}_R$ correction of $U_n/U_c$, as applied to the dimensionless tangential velocity  $U_t/U_c$ earlier (see (\ref{UtUcWR_1})),  to account for the precursor capillary wave effects, does not collapse the normal velocity data.  To address this scaling problem, we now analyse the average velocity of shrinkage of the cavity in two mutually perpendicular directions, \emph{viz.}, (i) in the equatorial plane in the horizontal direction, where the local radius of the cavity from its vertical axis of symmetry is maximum, and (ii) in the meridional plane along the vertical axis of symmetry (see figure~\ref{DeDnb_labelled}). 

\subsubsection{Average normal shrinkage velocity in the equatorial plane  and along the vertical axis}\label{shrinking_We_e}
\begin{figure}
    \centering
         \begin{subfigure}{0.47\textwidth}
        \includegraphics[width=\textwidth]{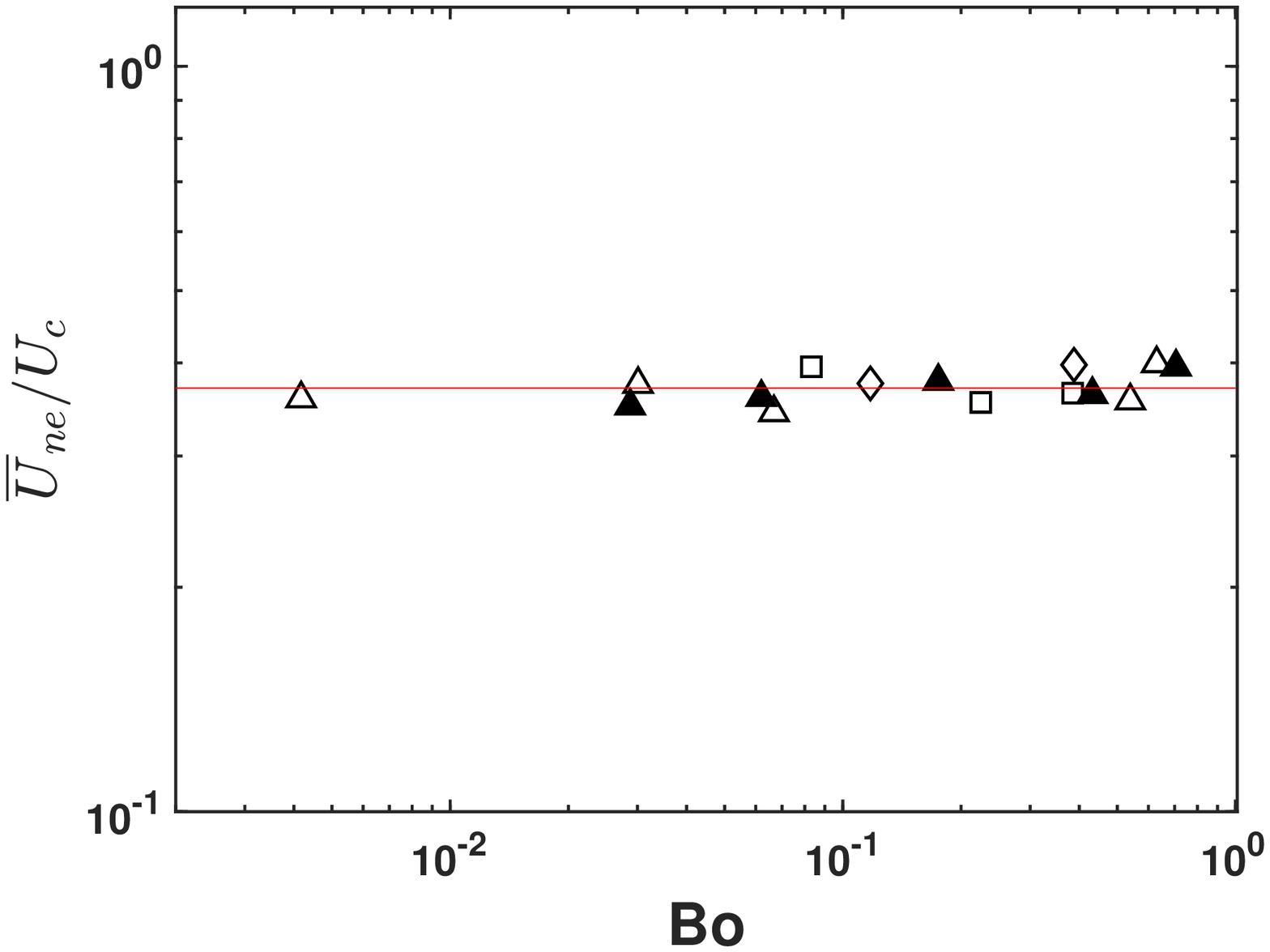}
        \caption{}
        \label{We_equitorial}
    \end{subfigure}
    \begin{subfigure}{0.51\textwidth}
        \includegraphics[width=\textwidth]{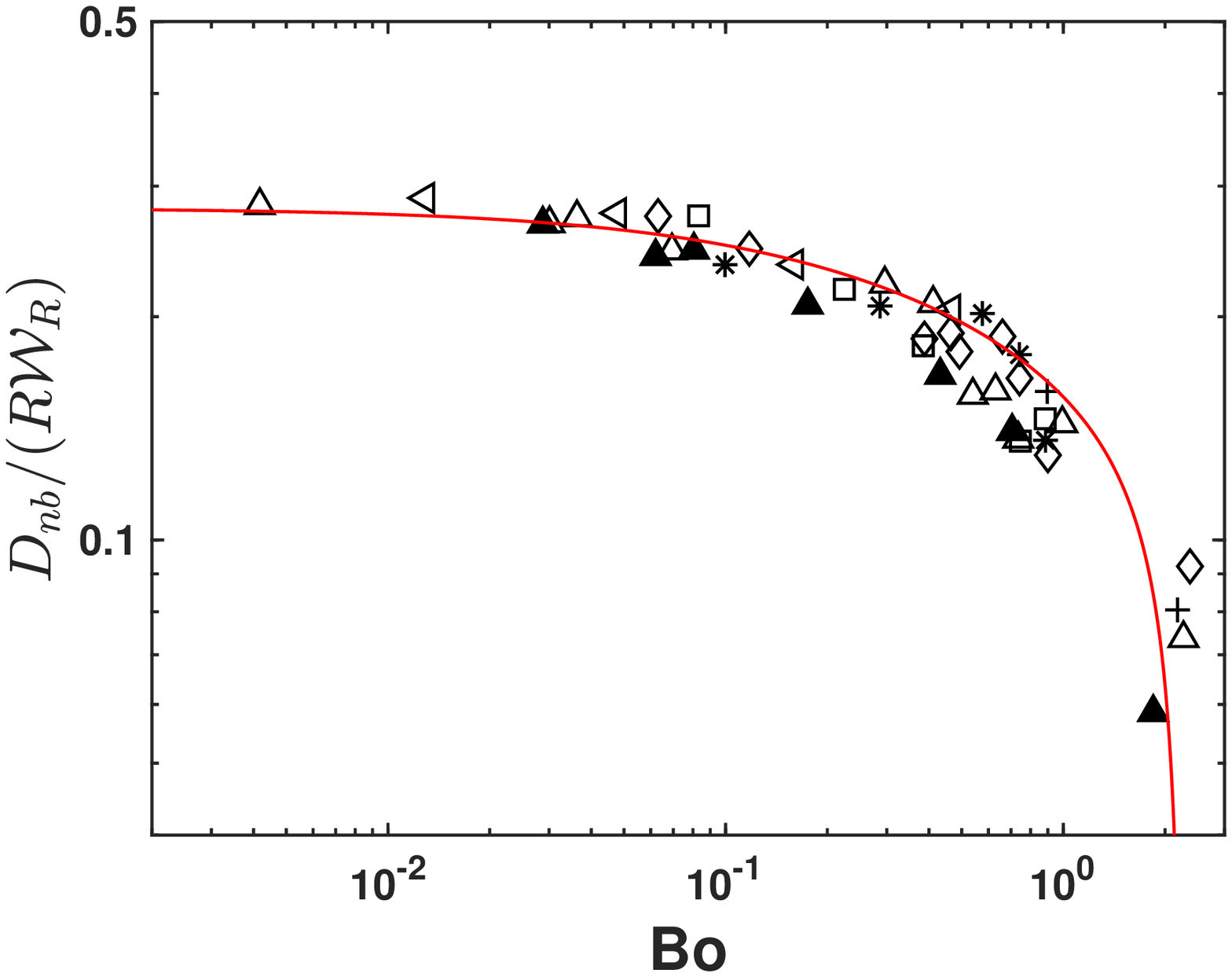}
        \caption{}
        \label{We_normal}
    \end{subfigure}
     \caption{(a), Dimensionless average shrinkage velocity   at the  equatorial plane (${\overline{U}}_{ne}/U_c$) verses  $Bo$.  --- , ${\overline{U}}_{ne}/U_c~=~0.37$. (b), dimensionless shrinkage of the  cavity bottom, ${D_{nb}}/(R{\mathcal{W}}_R)$, plotted as a function of $Bo$. ---, ${D_{nb}}/(R{\mathcal{W}}_R)= 0.14(Z_{c}/R)$. $\triangle$, Water; $\blacktriangle$, GW48 (30$^{\circ}$C); $\square$, GW55; $*$, GW68; $\lozenge$, GW72; $+$, 2-propanol; $\triangleleft$, Ethanol.}
    \label{}
\end{figure}
 Figure~\ref{We_equitorial} shows the average shrinkage velocity normal to the cavity boundary in the equatorial plane $ {\overline{U}}_{ne} = D_{ne}/t_{e}$, nondimensionalised by the capillary velocity scale,  as a function of $Bo$. Here, $D_{ne}$ is the average normal  displacement in the equatorial plane and  $t_e$ is the time taken for the kink to reach the equatorial plane; these are shown in figure~\ref{DeDnb_labelled}. 
 It is seen in figure~\ref{We_equitorial} that $ {\overline{U}}_{ne}/U_c$ is independent of $Bo$ and scales as
 \begin{equation}
\label{We_e_Oh_eqn}
 {\overline{U}}_{ne}=0.35~{U_c}.  
\end{equation} 
Thus, in the equatorial plane, where there is symmetry in the azimuthal direction, the velocity of shrinking scales with the capillary velocity $U_c$, devoid of any viscous and gravity effects.  

On the contrary, the normal shrinkage at the cavity bottom is strongly  dependent on $Bo$, as is seen in figure~\ref{We_normal}, where the dimensionless normal distance of shrinking at the bottom of the cavity  $D_{nb}/R$, corrected by ${\mathcal{W}}_R$, similar to that in  figure~\ref{Vt_inst1}, is plotted as a function of $Bo$.   The variation of the data in the figure is well represented by  the variation of the normalised cavity depth $Z_c/R$ with $Bo$, obtained using the closed form solution for $Z_c/R$ in terms of $Bo$, given in \cite{puthenveettil2018shape};  this representation reveals the dependence of $D_{nb}$ on $Z_c$.  Thus, the expression of the best fit of the data in figure~\ref{We_normal} is   
\begin{equation}
\label{Dn_Zc_expression1}
{D_{nb}}/(R{\mathcal{W}}_R)= 0.14(Z_{c}/R).
\end{equation} 
The velocity of cavity shrinking at the bottom is $ {\overline{U}}_{nb} = D_{nb}/t_{bc}$. Using (\ref{Dn_Zc_expression1})  and  $t_{bc}$ from (\ref{tbcRreqn}), the  average vertical shrinking rate of the cavity bottom then becomes,
\begin{equation}
\label{Unavg_Zcreln1}
  \frac{\overline{U}_{nb}}{U_c} = \frac{Z_c}{R}  \frac{\mathcal{W}_R} {\mathscr{L}}, 
\end{equation}
where $Z_c/R$ is the aspect ratio of the cavity shape.  Equation~(\ref{Unavg_Zcreln1})  shows that  the  precursory capillary waves act as deformations on the cavity surface, reducing cavity shrinking velocity at the bottom of the cavity, when  $Oh<0.02$. This reduced shrinking velocity of the cavity in low viscosity fluids can also be seen in figure~\ref{Vn_curvefit}. RHS of  (\ref{Unavg_Zcreln1}) tends to a constant. The resulting capillary velocity scaling of $\overline{U}_{nb}$ is in agreement with the radial  velocity scaling of the cavity  in the inviscid limit  proposed by  \cite{Ganan-calvo_lopez-herrera_2021} for $Oh\ll0.04$. However, the present  $\overline{U}_{nb}$   differs from the viscous-capillary velocity ($V_{\mu}$) dependence of radial velocity proposed by  \cite{Ganan-calvo_lopez-herrera_2021}, for the final stage of collapse at around $Oh=0.04$. This difference is possibly due to the fact that $\overline{U}_{nb}$ is  the velocity averaged over $t_{bc}$, which may not capture the sharp changes in velocities near  the flow focusing.  

In any case it is clear from (\ref{Unavg_Zcreln1}) and (\ref{wave_resi_datamatch}) is that the bottom region of the cavity, prior to the flow focusing, closes with two different velocity scales, originating from two different physical mechanisms. The symmetric interface velocity assumption used by \cite{Ganan-calvo_lopez-herrera_2021} or a purely horizontal converging flow proposed by \cite{gordillo2019capillary} would then need modifications in light of this understanding. It is also of interest to note that the cavity shrinking velocity at the equatorial plane ${\overline{U}}_{ne}$ (\ref{We_e_Oh_eqn}) is independent of $Oh$.  This is in contrast with that of $ {\overline{U}}_{nb}$ (\ref{Unavg_Zcreln1}), which has a dependence on viscosity through ${\mathcal{W}_R}$, the wave resistance factor due to the presence of precursory  capillary waves. This difference is expected to be because the precursory capillary waves affect the  cavity collapse only after the kink crosses the equatorial plane. In the final phase of cavity collapse, i.e. at the arrival of the kink at the cavity bottom,  the  rise in the normal velocity in low viscosity fluids is almost an order of magnitude higher (see figure~\ref{Vn_curvefit}) than in fluids of high viscosity. This rise in $U_n$ occurs when the capillary waves  converge at the cavity bottom.

\section{The cavity filling rate}\label{massfilltext}
\begin{figure}
    \centering
    \begin{subfigure}{0.40\textwidth}
        \includegraphics[width=\textwidth]{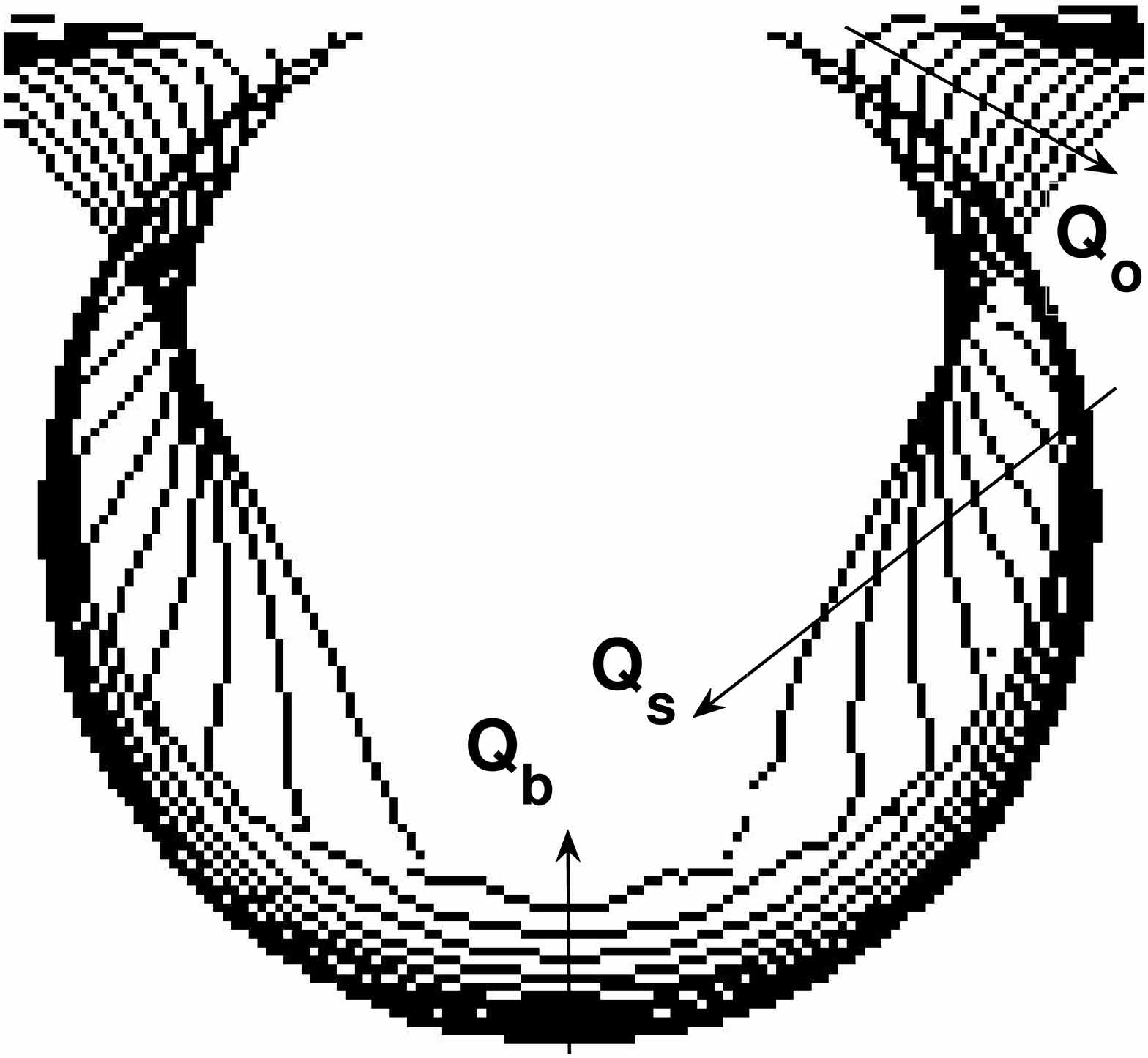}
        \caption{}
        \label{R175contoursuperimpos}
    \end{subfigure}
    \begin{subfigure}{0.53\textwidth}
        \includegraphics[width=\textwidth]{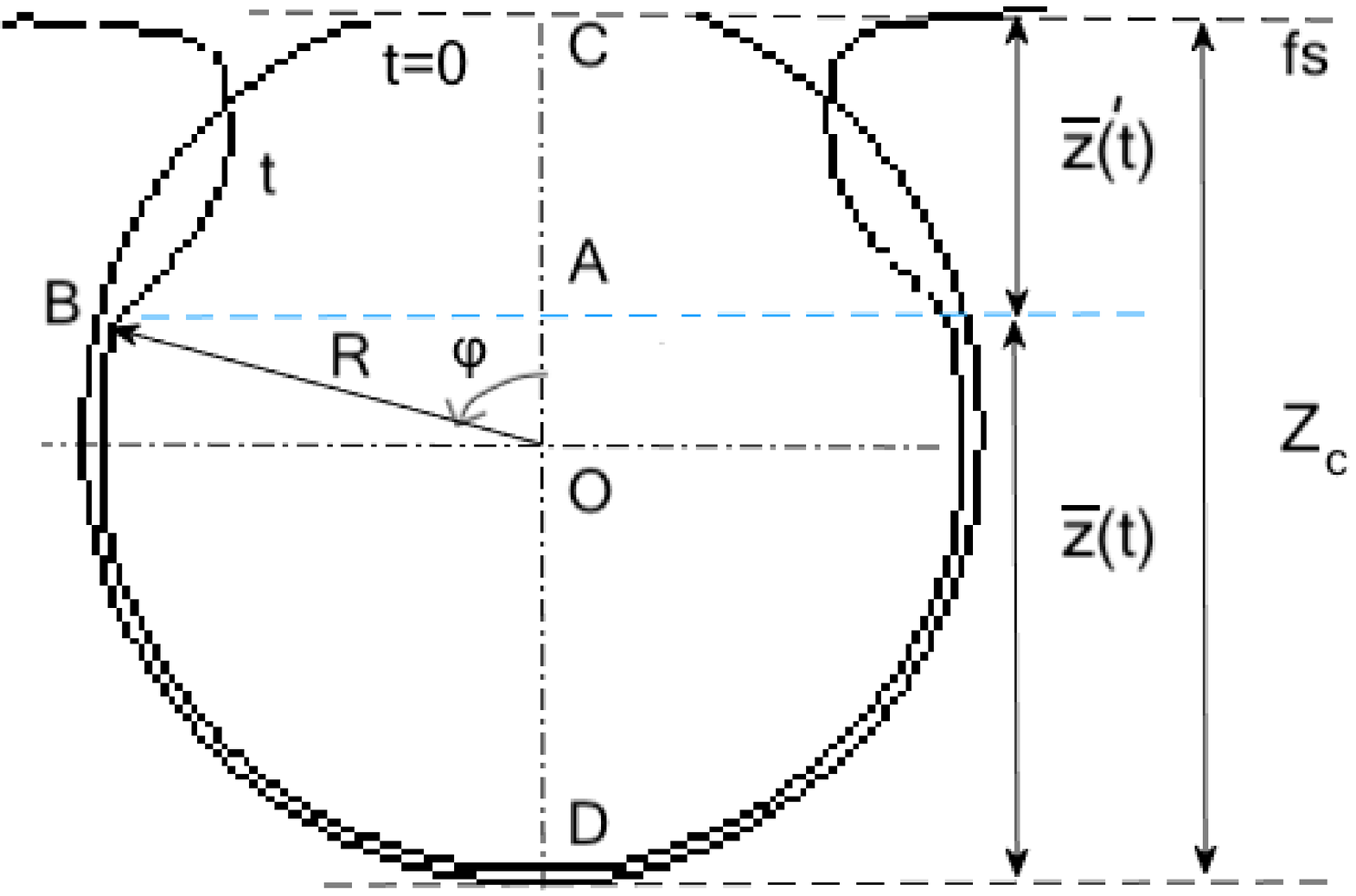}
        \caption{}
        \label{theta_description}
    \end{subfigure}
     \caption{(a) Superimposed contours of collapsing cavity at different time instances of bubble $R=0.175$mm in water, starting from the static shape at $t=0$ until the conical cavity shape is reached. The successive contours are separated by $10\mu$s. Movie~3. (b), Contours of initial static cavity configuration, at $t=0$, and at a later time $t>0$.  The figure shows various parameters related with the moving kink; the free-surface is indicated by fs.}
    \label{cavitycollapse_representation}
\end{figure}
\begin{figure}
  \centering
  \includegraphics[width=0.95\textwidth]{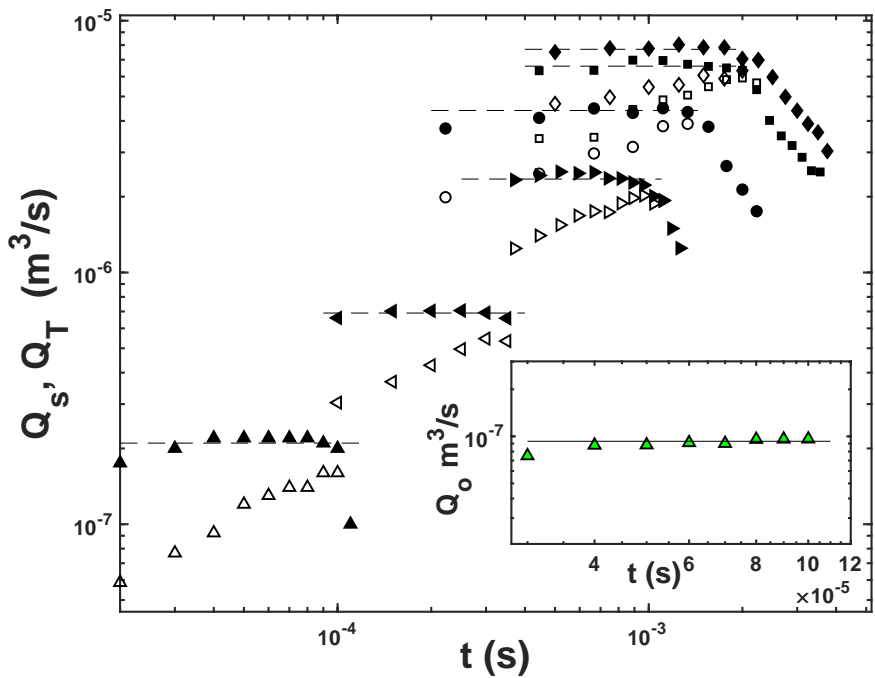}
    \caption{The rate of  filling of the cavity from the tangential direction, $Q_s$ (hollow symbols), and the total filling rate $Q_T$ (solid symbols) as functions  of time, for bubbles of different $Bo$ and $Oh$. $\lozenge$, $R=2.15$mm ($Bo=0.63$, $Oh=0.00255$); $\square$, $R=1.89$mm ($Bo=0.49$, $Oh=0.0027$); $\bigcirc$,  $R=1.47$mm ($Bo=0.3$, $Oh=0.0031$);  $\triangleleft$, $R=0.47$mm ($Bo=0.03$, $Oh=0.0055$); $\triangle$, $R=0.175$mm ($Bo=0.0042$, $Oh=0.0099$) (all data in water). $\triangleright$, $R=1.04$mm ($Bo=0.17$, $Oh=0.0139$) in GW48.  The volume out-flux, $Q_o$, measured for $R=0.175$mm in water, is shown in the inset.}
  \label{dimensionalQ}
\end{figure}
Three volume fluxes can be identified in relation with the cavity boundary movement, as indicated in figure~\ref{QtQn_labelled}, and measured, as discussed in \S~\ref{expsetup}. Figure~\ref{R175contoursuperimpos} shows the contours of the collapsing cavity at different time instances, starting from the static shape at $t=0$ until the conical cavity shape is reached, of a bubble of radius  $R=0.175$mm in water.  The volume fluxes that can be identified from the figure are, (i)  the side (tangential) volume influx $Q_s$, (ii) the bottom (normal) influx $Q_b$ and (iii) the side volume out-flux $Q_o$, with the total filling rate being $Q_T=Q_s+Q_b$. Figure~\ref{dimensionalQ} shows the variation of the side volume influx $Q_s$ (denoted by hollow symbols) and the total volume influx $Q_T$  (solid symbols) as a function of the time, for different size bubbles, in water and GW48. The horizontal and vertical axes in   figure~\ref{dimensionalQ} span three orders of magnitude of time and volume flux, respectively. Each data set of $Q_{T}$  shows an approximately  constant value with time, indicated by a horizontal dashed line, and then suddenly drops off. This sudden change in $Q_T$ is indicative of an unaccounted volume out-flux, due to the creation of the jet inside of the cavity, coinciding with the conical cavity shape. The side volume flux $Q_s$ which is initially a small fraction of $Q_T$, increases with time and represents nearly the total volume flux when the cavity becomes a cone. The difference between $Q_T$ and $Q_s$ in  figure~\ref{dimensionalQ} corresponds to the bottom influx $Q_b$ at any given  time. The inset in figure~\ref{dimensionalQ} shows the  variation of the volume out-flux $Q_o$ with time, measured for a bubble of radius  $R=0.175$mm in water. The volume out-flux $Q_o$ is not entirely negligible. However, since it is practically constant in time and since $Q_T$ is constant, $Q_{T}-Q_{o}$ is also a constant so that the volume expansion of the cavity at the free-surface ($Q_o$) will not change the functional behaviour of $Q_T$ with time. Therefore,  $Q_o$ is not considered further  in our analysis.

\begin{figure}
  \centering
  \includegraphics[width=0.96\textwidth]{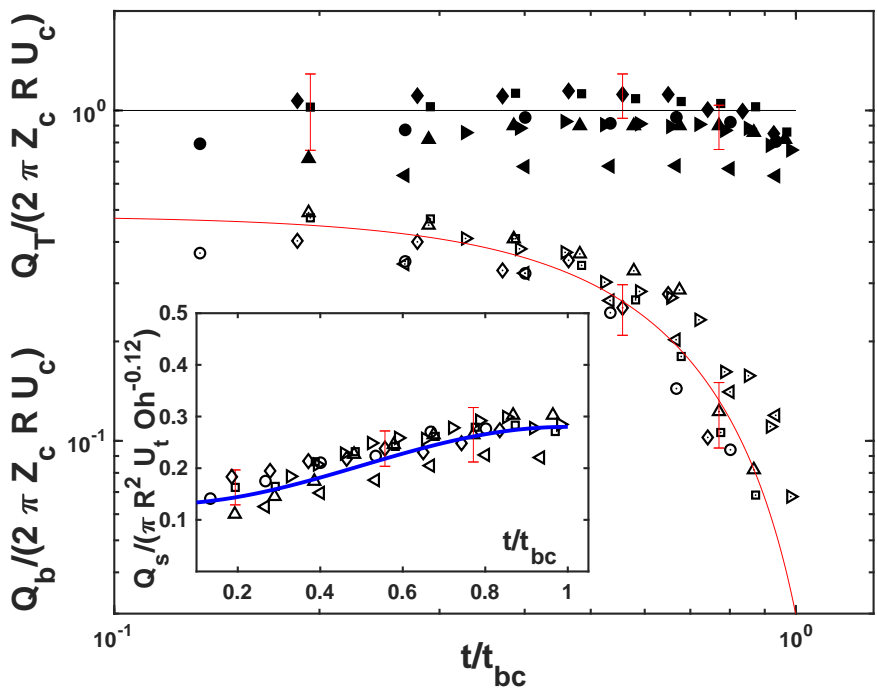}
    \caption{Variation of the normalised  total volume influx $Q_{T}/2\pi R Z_c U_c$ and the non-dimensional volume influx in the (bottom)  normal direction $Q_b/2\pi Z_c R U_c$  as a function of the  normalised time $t/t_{bc}$.  In the inset: volume influx along the tangential direction $Q_s$ is nondimensionalised with $\pi R^2 U_tOh^{-0.12}$ and plotted against the non-dimensional time $t/t_{bc}$. --- (black),  $Q_{T}/2\pi R Z_c U_c=1$. --- (red), $Q_b(t)/2\pi Z_c R U_c = 0.5\cos^2 \left(0.42\pi{t}/{t_{bc}}\right)$.  In the inset:  --- (blue), $Q_s(t)/\pi R^2 U_t Oh^{-0.12} = 0.13 + 0.15\sin^2 \left(0.5\pi{t}/{t_{bc}}\right)$.  Symbols with dots represent volume influx in the (bottom)  normal direction $Q_b$. The rest of the symbols are the same as in figure~\ref{dimensionalQ}.}
  \label{nondimensionalQ}
\end{figure}

We now obtain a scaling for $Q_s$ as follows. The characteristic area of the side flux is $2\pi R \overline{z}'(t)$, where $\overline{z}'(t)$ is  the height above the kink up to the free-surface level (see figure~\ref{theta_description}),   with the velocity being the tangential velocity of the kink $U_t$ (\S~\ref{Uttext}). Although the kink moves with a velocity  $U_t$, the upper interior region of the concave boundary lags behind the kink (see A in figure~\ref{cavitycollapse_labelled}). Such a velocity difference inside the side boundary implies the existence of a shear region, which needs to be accounted for in the side volume flux. Hence,  we include a viscous correction term of the form $Oh^d$ to estimate the side flux as $Q_s\sim 2\pi R  U_t Oh^d \overline{z}'(t) \sim 2\pi R U_t Oh^d (R-R\cos \phi)$, which then yields, 
 \begin{equation}
     \frac{Q_s}{\pi R^2 U_t Oh^d}\approx \mathsf{a_1} \sin^2 (\phi/2),
     \label{Qs_nondim_1}
 \end{equation}
 where $\mathsf{a_1}$ is a constant to be determined from experiments. In (\ref{Qs_nondim_1}) $\phi={\omega t}+\phi_0$ is the phase angle of the moving kink (see figure~\ref{theta_description}),    where $\omega$ is the circular frequency and $\phi_0\approx\theta$ (see figure~\ref{DeDnb_labelled}) is the phase angle when the cavity opens.  In the inset of figure~\ref{nondimensionalQ} the dimensionless side volume flux $Q_s/\pi R^2 U_t Oh^{-0.12}$ is plotted against the dimensionless time $t/t_{bc}$.
 Equation~(\ref{Qs_nondim_1})  collapses the data, with the final expression based on the data fit being,  
\begin{equation}
\label{Qs_nondim_2}
\frac{Q_s (t)}{\pi R^2 U_t Oh^{-0.12}} ~ = ~ 0.12 + 0.15 \sin^2 \left(\frac{0.5\pi t}{t_{bc}}\right),
\end{equation}
 where the first term on the right-hand side of (\ref{Qs_nondim_2})  is the initial side flux.

We now consider the scaling of the bottom flux. The characteristic area below the kink is $2\pi R \overline{z}(t)$, where $\overline{z}(t)$ is the vertical distance between the kink and the cavity bottom (see figure~\ref{theta_description}). The velocity of shrinking in the equatorial plane follows the capillary velocity  (see (\ref{We_e_Oh_eqn})).  Hence, the  normal influx below the kink  is   
 $Q_b\sim 2\pi R \overline{z}(t) U_c \sim 2\pi R U_c (Z_c-R(1-\cos \phi))$.  Taking $2R/Z_c\approx 1$ for small $Bo$ \citep{puthenveettil2018shape} results in the simplified relation,  
 \begin{equation}
     \frac{Q_b}{2\pi R Z_c U_c}\approx \mathsf{a_2} \cos^2 (\phi/2), 
     \label{Qb_final_reln}
 \end{equation}
 where $\mathsf{a_2}$ is a numerical prefactor. 
Figure~\ref{nondimensionalQ} shows that (\ref{Qb_final_reln}) collapses the $Q_b$ data at various $Bo$ and $Oh$, with the best fit relation for the  nondimensional bottom flux, $Q_b(t)/2\pi R Z_c U_c$, shown in figure~\ref{nondimensionalQ} being
\begin{equation}
\label{Qb_nondim2}
 \frac{Q_b (t)}{2\pi R Z_c U_c} ~ = ~ 0.5 \cos^2 \left(\frac{0.42\pi t}{t_{bc}}\right).
\end{equation}
The phase angle of $Q_s$ in (\ref{Qs_nondim_2}) $\pi t$ is slightly larger than that of $Q_b$ in (\ref{Qb_nondim2}), $0.84\pi t$, an artefact of the side flux area being at the top of the bottom flux area.

We now assume that the retracting rim solely creates a wave-like propagating disturbance without causing any effective mass transfer down the cavity.  Then, the total cavity filling rate  $Q_T\approx (Q_b+Q_s)|_{t\rightarrow 0}\approx 2\pi R Z_c U_c $. This means that the initial side flux in (\ref{Qs_nondim_2}) is then due to the normal shrinking. Figure~\ref{nondimensionalQ} shows that $Q_{T}/2\pi R Z_c U_c$  collapses the data reasonably well. From the plot, the equation of best fit is 
\begin{equation}
\label{totalQfinal}
 \frac{Q_{T} (t)}{2\pi R Z_c U_c} ~ \approx ~ 1, 
\end{equation}
validating our assumptions to arrive at the above relation. 
  
 Equation~(\ref{totalQfinal}) shows that the total volume flux is entirely due to the normal shrinkage velocity of the cavity. When $Bo<0.1$, the normalised cavity depth  $Z_c/R\rightarrow2$, and is independent of $Bo$ so that gravity effects in $Q_T$ becomes negligible. Gravity effects become significant in $Q_T$ through the $Bo$ dependency of $Z_c$, in the moderate to large bubble size range ($Bo>0.1$)
 \citep{krishnan_hopfinger_puthenveettil_2017,puthenveettil2018shape}.  Towards $t=t_{bc}$, when the cavity becomes conical,  the total mass flux $\rho Q_T$ initiates a jet by momentum exchange. More details  on flow focusing and jetting are provided in Appendix~\ref{mtm_model_jet}.

\section{Conclusions}\label{conclusions}
Following the  disintegration of the thin film at the top of a floating bubble, the rim retraction leads to the formation of a kink (intersection of the concave with the convex cavity boundary) that travels tangentially along the cavity boundary, with  capillary waves, absent in high viscosity fluids, moving ahead of the kink. Simultaneously,  the cavity shrinks due to the sudden gas pressure reduction after film rupture. These two different mechanisms lead to the tangential ($U_t$) and normal velocities ($U_n$) of the collapsing cavity boundary.  The  tangential motion of the kink, combined with overall inward cavity shrinkage due to gas pressure reduction is a unique feature of surface bubble cavity collapse, not encountered in open  cavity collapse problems as treated by \cite{zeff2000singularity,Bergmann2006PRL,Bartolo_dropimpact_06PRL,duclaux2007,Das_Hopfinger_Faradaywaves2008,benusiglio2014,thoroddsen_takehara_nguyen_etoh_2018,yang_tian_thoroddsen_2020,krishnan2022impact}.  
   
Surface bubble cavity collapse is  a capillary driven phenomenon, with viscosity and gravity affecting the collapse dynamics by, respectively, damping of precursory capillary waves and by a reduction of the static cavity depth ($Z_c$). An increase in fluid viscosity increases the tangential and normal velocities of the kink because of the progressive damping of precursor capillary waves. Using an energy model of the kink region, we show that  ${U_{t}}\approx4.5~U_c~{\mathcal{W}}_R$ (\ref{wave_resi_datamatch}), where $U_c$ is the capillary velocity, ${\mathcal{W}}_R= { (1-\sqrt{Oh{\mathscr{L}}})^{-1/2}}$ (\ref{WREqn}) is the wave resistance factor with $\mathscr{L}(Bo)=\pi -{R_r}/{R}$ (\ref{LBo_def}) being the correction for the path length of the travel of the kink, $Oh$, the Ohnesorge number, $Bo$, the Bond number and $R_r$, the rim radius.

The sudden release of compressed gas from the bubble cavity, immediately after the thin film rupture, causes an overall inward shrinking of the cavity. It produces a normal velocity component to the kink. In the horizontal  equatorial plane, the normal kink velocity scales with  $U_c$ (\ref{We_e_Oh_eqn}), devoid of viscous and gravity effects. In contrast, the bottom part of the cavity shrinks vertically upwards with a velocity scale ${U_c (Z_c/R) {\mathcal{W}}_R}{{\mathscr{L}}}^{-1}$ (\ref{Unavg_Zcreln1}). The viscous effect in the vertical shrinkage at the bottom, ${\mathcal{W}}_R$, is due to the deformations by the  precursor capillary waves on the cavity. The gravity effect on this shrinkage originates from the initial static geometry of the cavity, indicated by the aspect ratio of the cavity $Z_c/R$ and the path correction  $\mathscr{L}$.

The total time of cavity collapse is shown to scale as  ${t_{bc}}\approx 0.13 t_c{\mathscr{L}}$ (\ref{tbcRreqn}) with the gravity dependency being due to the $Bo$-dependence of the kink trajectory ${\mathscr{L}}$. This leads to the understanding that the damping of the precursory capillary waves follows a modified relation  $\lambda/R\approx{Oh^{1/2}{\mathscr{L}}^{1/2}}$ (\ref{lambda_Oh_Bo_new}) indicating that the damping is slightly reduced with an increase in Bond number.

The mass flux of cavity filling $\rho Q_T$ consists of a sum of the side flux $\rho Q_s$ and normal bottom flux $\rho Q_b$. We show that $Q_T\approx 2\pi RZ_c U_c$ (\ref{totalQfinal}), which then depends on the aspect ratio of the cavity $Z_c/R$,  which is a function of $Bo$. The entire magnitude of $\rho Q_T$ originates from the normal shrinkage of the cavity, showing that the kink movement, initiated by the rim retraction, is similar to the propagation of a wave front with no effective mass transfer. Indeed, the  tangential velocity $U_t$, which is constant with respect to time for a given bubble, corresponds closely to the phase velocity of a capillary wave of wavelength $\lambda$, i.e. $U_t\approx{c_p}$ (\ref{cp_eqn}), where $c_p=(2\pi)^{1/2}\sqrt{{\sigma}/{\rho \lambda}}$. With the experimental value $\lambda/R\approx {0.36}$, we get $c_p\approx4.2U_c$, which is close to $U_t$  (\ref{wave_resi_datamatch}), neglecting the weak dependency on viscosity and gravity. 

At the bottom of the cavity, there is an exchange of momentum, via pressure build-up,  due to the mass flux $\rho Q_T$ of the cavity with the initial jet mass flux $\rho\pi{r_j}^2U_j$, where  $r_j\sim r_b$ is the  conical cavity base radius and $U_j$ the jet velocity (see Appendix~\ref{mtm_model_jet}). This exchange gives a jet Weber number scaling $(U_j/U_c)^2=C^2(Z_cR/{r_j}^2)^2$, where the coefficient $C\approx0.5$ because the momentum exchange is not perfect. From experiments  $R/r_j\approx3$ so that the jet Weber number is $We_j\approx 350$ in the limit of $Bo<0.1$.     

\medskip

\noindent \textbf{Funding.} The authors would like to acknowledge the partial financial support from DST, Government of India, under the FIST Grants SR/FST/ETII-017/2003, SR/FST/ETII-064/2015 and the Core Research Grants SR/S3/MERC/028/2009, CRG/2021/007497 for this study.

\medskip

\noindent \textbf{Declaration of interests.} The authors report no conflict of interest.

\appendix 
\renewcommand\thefigure{\thesection.\arabic{figure}} 

\section{Momentum balance at the cavity bottom}\label{mtm_model_jet}
 \cite{DeikePRF2018} and   \cite{duchemin2002jet} observed in their numerical simulations, two successive velocity (or pressure) peaks at the cavity bottom, with the second peak being the highest. Considering the initial peak to be due to the capillary waves, and the second one to be due to the kink, the momentum of the precursor waves is then not significant, compared with that of  the kink, an aspect clarified by \cite{Alfonso_2018_PhysRevFluids}. However, the volume flux (or mass flux) is associated with the capillary velocity $U_c$ due to the normal shrinking of the cavity (see (\ref{totalQfinal})), while $U_t$ is only a propagation velocity, carrying little mass.  
The rate of change of momentum of the liquid of the collapsing cavity is then ${d}(mU_c)/{dt}$, where $m$ is the associated liquid mass.  Differentiation gives  $\dot{m}U_c$+$m\dot{U_c}$, where $\dot{m}$ is the mass flux of the cavity collapse.   Since $U_c$ is a constant with respect to time, we get,
\begin{equation}
\frac{d}{dt}(mU_c)=\dot{m}U_c.
\label{mtmfluxeq1}
\end{equation}

The momentum flux in (\ref{mtmfluxeq1}) will appear as a force during the axisymmetric flow-focusing at the cavity bottom. The corresponding pressure build-up  is
\begin{equation}
\label{p_buildup}
p\simeq{\dot{m}U_c}/2\pi {r_b}^2,
\end{equation}
 where $2\pi {r_b}^2$ is the characteristic area at the base  of the conical cavity of bottom radius $r_b$ (see figure~\ref{DeDnb_labelled}). By substituting the total mass influx $\dot{m}=\rho Q_T$ from  (\ref{totalQfinal}) in (\ref{p_buildup}), we get an estimate of the pressure at the bottom  as, 
	\begin{equation}
	p\simeq \left(\frac{Z_c}{R}\right){\rho {U_c}^2}\left(\frac{R}{r_b}\right)^2.
	\label{p_expression}
	\end{equation}
Equation (\ref{p_expression}) shows that the pressure build-up at the bottom of the cavity, has primarily a capillary-inertial scaling, of the form $\rho {U_c}^2$. As the aspect ratio of the cavity ${Z_c}/{R}$  is a function of $Bo$, as given by \cite{puthenveettil2018shape}, $p$ also depends on $Bo$. Similarly, the effect of precursory waves on the pressure is accounted for by the term  $(R/r_b)^2$ in (\ref{p_expression}).  Note that  a reduced area of impact ($\sim{{r_b}^2}$) in the absence of capillary waves increases the impact pressure.

The pressure impulse of the impact $P=\int_{t_i}^{t_f} p~dt$,  where the  subscripts $i$ and the $f$ denote the initial and final values of the  time $t$,  is estimated to be  
\begin{equation}
\label{P_deltat}
    P\approx p\Delta t,
\end{equation}
where $\Delta t=t_f -t_i$ is the time scale of the impact. The natural choice of the time scale of impact in this capillary-driven flow focusing is 
 \begin{equation}
 \label{natural_timescale_impact}
    \Delta t\simeq {r_b}/{U_c}. 
 \end{equation}
As the gradient of pressure impulse  drives the jet in the axial direction, 
\begin{equation}
\label{Uj_PI}
    U_j=-\nabla ({P}/{\rho}).
\end{equation}
 Substituting (\ref{p_expression}) and (\ref{natural_timescale_impact}) in (\ref{P_deltat}) and the resulting expression for $P$ in (\ref{Uj_PI}), we obtain,
\begin{equation}
U_j \simeq~\nabla\left(\frac{ R~ Z_c~ {U_c}}{r_b}\right).
\label{uaubuj}
\end{equation}
Approximating the gradient operator  $\nabla$ as $ 1/r_b$,  gives the jet Weber number $We_j$ as 
\begin{equation}
   We_j = \left(\frac{U_j}{U_c}\right)^2\sim \left(\frac{Z_c}{R}\right)^2 \left(\frac{R}{r_b}\right)^4,
\end{equation}
 the same scaling as that proposed by \cite{krishnan_hopfinger_puthenveettil_2017}. The jet radius $r_j\sim r_b$ \citep{Ganan-calvo_lopez-herrera_2021} with the ratio $R/r_b$ depending on the presence, or not, of the precursor capillary waves.

\bibliographystyle{jfm}
\bibliography{Revised_Cavity_collapse_paper}

\begin{thebibliography}{43}
\expandafter\ifx\csname natexlab\endcsname\relax\def\natexlab#1{#1}\fi
\def\au#1{#1} \def\ed#1{#1} \def\yr#1{#1}\def\at#1{#1}\def\jt#1{\textit{#1}}
  \def\bt#1{#1}\def\bvol#1{\textbf{#1}} \def\vol#1{#1} \def\pg#1{#1}
  \def\publ#1{#1}\def\arxiv#1{#1}\def\org#1{#1}\def\st#1{\textit{#1}}

\bibitem[Bartolo {\em et~al.\/}(2006)Bartolo, Josserand \&
  Bonn]{Bartolo_dropimpact_06PRL}
{\sc \au{Bartolo, D.}, \au{Josserand, C.} \& \au{Bonn, D.}} \yr{2006}
  \at{Singular jets and bubbles in drop impact}.  \jt{Phys. Rev. Lett.}
  \bvol{96},  \pg{124501}.

\bibitem[Benusiglio {\em et~al.\/}(2014)Benusiglio, Qu{\' e}r{\' e} \&
  Clanet]{benusiglio2014}
{\sc \au{Benusiglio, A.}, \au{Qu{\' e}r{\' e}, D.} \& \au{Clanet, C.}}
  \yr{2014}  \at{Explosions at the water surface}.  \jt{Journal of Fluid
  Mechanics}  \bvol{752},  \pg{123--139}.

\bibitem[Bergmann {\em et~al.\/}(2006)Bergmann, van~der Meer, Stijnman,
  Sandtke, Prosperetti \& Lohse]{Bergmann2006PRL}
{\sc \au{Bergmann, R.}, \au{van~der Meer, D.}, \au{Stijnman, M.}, \au{Sandtke,
  M.}, \au{Prosperetti, A.} \& \au{Lohse, D.}} \yr{2006}  \at{Giant bubble
  pinch-off}.  \jt{Phys. Rev. Lett.}  \bvol{96},  \pg{154505}.

\bibitem[Blanchard(1963)]{blanchard1963electrification}
{\sc \au{Blanchard, D.~C.}} \yr{1963}  \at{The electrification of the
  atmosphere by particles from bubbles in the sea}.  \jt{Progress in
  oceanography}  \bvol{1},  \pg{73IN7113--112202}.

\bibitem[Blanco–Rodríguez \& Gordillo(2021)]{blanco-rodriguez_gordillo_2021}
{\sc \au{Blanco–Rodríguez, F.~J.} \& \au{Gordillo, J.~M.}} \yr{2021}  \at{On
  the jets produced by drops impacting a deep liquid pool and by bursting
  bubbles}.  \jt{Journal of Fluid Mechanics}  \bvol{916},  \pg{A37}.

\bibitem[Boulton-Stone \& Blake(1993)]{BSB}
{\sc \au{Boulton-Stone, J.~M.} \& \au{Blake, J.~R.}} \yr{1993}  \at{Gas bubbles
  bursting at a free surface}.  \jt{Journal of Fluid Mechanics}  \bvol{254},
  \pg{437--466}.

\bibitem[Brasz {\em et~al.\/}(2018)Brasz, Bartlett, Walls, Flynn, Yu \&
  Bird]{Brasz_PhysRevFluids2018}
{\sc \au{Brasz, C.~F.}, \au{Bartlett, C.~T.}, \au{Walls, P. L.~L.}, \au{Flynn,
  E.~G.}, \au{Yu, Y.~E.} \& \au{Bird, J.~C.}} \yr{2018}  \at{Minimum size for
  the top jet drop from a bursting bubble}.  \jt{Phys. Rev. Fluids}  \bvol{3},
  \pg{074001}.

\bibitem[Burton {\em et~al.\/}(2005)Burton, Waldrep \&
  Taborek]{burton2005scaling}
{\sc \au{Burton, J.~C.}, \au{Waldrep, R.} \& \au{Taborek, P.}} \yr{2005}
  \at{Scaling and instabilities in bubble pinch-off}.  \jt{Phys. Rev. Lett.}
  \bvol{94}~(18),  \pg{184502}.

\bibitem[Ga\~n\'an Calvo(2017)]{AlfonsoPRL2017}
{\sc \au{Ga\~n\'an Calvo, A.~M.}} \yr{2017}  \at{Revision of bubble bursting:
  Universal scaling laws of top jet drop size and speed}.  \jt{Phys. Rev.
  Lett.}  \bvol{119},  \pg{204502}.

\bibitem[Ga\~n\'an Calvo(2018)]{Alfonso_2018_PhysRevFluids}
{\sc \au{Ga\~n\'an Calvo, A.~M.}} \yr{2018}  \at{Scaling laws of top jet drop
  size and speed from bubble bursting including gravity and inviscid limit}.
  \jt{Phys. Rev. Fluids}  \bvol{3},  \pg{091601}.

\bibitem[Ga\~n\'an Calvo \&
  López-Herrera(2021)]{Ganan-calvo_lopez-herrera_2021}
{\sc \au{Ga\~n\'an Calvo, A.~M.} \& \au{López-Herrera, J.~M.}} \yr{2021}
  \at{On the physics of transient ejection from bubble bursting}.  \jt{Journal
  of Fluid Mechanics}  \bvol{929},  \pg{A12}.

\bibitem[Das \& Hopfinger(2008)]{Das_Hopfinger_Faradaywaves2008}
{\sc \au{Das, S.~P.} \& \au{Hopfinger, E.~J.}} \yr{2008}  \at{Parametrically
  forced gravity waves in a circular cylinder and finite-time singularity}.
  \jt{Journal of Fluid Mechanics}  \bvol{599},  \pg{205--228}.

\bibitem[Deike {\em et~al.\/}(2018)Deike, Ghabache, Liger-Belair, Das, Zaleski,
  Popinet \& S\'eon]{DeikePRF2018}
{\sc \au{Deike, L.}, \au{Ghabache, E.}, \au{Liger-Belair, G.}, \au{Das, A.~K.},
  \au{Zaleski, S.}, \au{Popinet, S.} \& \au{S\'eon, T.}} \yr{2018}
  \at{Dynamics of jets produced by bursting bubbles}.  \jt{Phys. Rev. Fluids}
  \bvol{3},  \pg{013603}.

\bibitem[Doshi {\em et~al.\/}(2003)Doshi, Cohen, Zhang, Siegel, Howell, Basaran
  \& Nagel]{Doshi_persistence_universality}
{\sc \au{Doshi, P.}, \au{Cohen, I.}, \au{Zhang, W.~W.}, \au{Siegel, M.},
  \au{Howell, P.}, \au{Basaran, O.~A.} \& \au{Nagel, S.~R.}} \yr{2003}
  \at{Persistence of memory in drop breakup: The breakdown of universality}.
  \jt{Science}  \bvol{302}~(5648),  \pg{1185--1188},  \arxiv{arXiv:
  http://science.sciencemag.org/content/302/5648/1185.full.pdf}.

\bibitem[Duchemin {\em et~al.\/}(2002)Duchemin, Popinet, Josserand \&
  Zaleski]{duchemin2002jet}
{\sc \au{Duchemin, L.}, \au{Popinet, S.}, \au{Josserand, C.} \& \au{Zaleski,
  S.}} \yr{2002}  \at{Jet formation in bubbles bursting at a free surface}.
  \jt{Physics of Fluids}  \bvol{14}~(9),  \pg{3000--3008}.

\bibitem[Duclaux {\em et~al.\/}(2007)Duclaux, Caille, Duez, Ybert, Bocquet \&
  Clanet]{duclaux2007}
{\sc \au{Duclaux, V.}, \au{Caille, F.}, \au{Duez, C.}, \au{Ybert, C.},
  \au{Bocquet, L.} \& \au{Clanet, C.}} \yr{2007}  \at{Dynamics of transient
  cavities}.  \jt{Journal of Fluid Mechanics}  \bvol{591},  \pg{1--19}.

\bibitem[Feng {\em et~al.\/}(2014)Feng, Roch{\'e}, Vigolo, Arnaudov, Stoyanov,
  Gurkov, Tsutsumanova \& Stone]{feng2014nanoemulsions}
{\sc \au{Feng, J.}, \au{Roch{\'e}, M.}, \au{Vigolo, D.}, \au{Arnaudov, L.~N.},
  \au{Stoyanov, S.~D.}, \au{Gurkov, T.~D.}, \au{Tsutsumanova, G.~G.} \&
  \au{Stone, H.~A.}} \yr{2014}  \at{Nanoemulsions obtained via bubble-bursting
  at a compound interface}.  \jt{Nature physics}  \bvol{10}~(8),
  \pg{606--612}.

\bibitem[Ghabache {\em et~al.\/}(2014)Ghabache, Antkowiak, Josserand \&
  S{\'e}on]{ghabache2014physics}
{\sc \au{Ghabache, E.}, \au{Antkowiak, A.}, \au{Josserand, C.} \& \au{S{\'e}on,
  T.}} \yr{2014}  \at{On the physics of fizziness: How bubble bursting controls
  droplets ejection}.  \jt{Physics of Fluids (1994-present)}  \bvol{26}~(12),
  \pg{121701}.

\bibitem[Gordillo \&
  Rodr{\'\i}guez-Rodr{\'\i}guez(2019)]{gordillo2019capillary}
{\sc \au{Gordillo, J.~M.} \& \au{Rodr{\'\i}guez-Rodr{\'\i}guez, J.}} \yr{2019}
  \at{Capillary waves control the ejection of bubble bursting jets}.
  \jt{Journal of Fluid Mechanics}  \bvol{867},  \pg{556--571}.

\bibitem[Ismail {\em et~al.\/}(2018)Ismail, Ga\~n\'an Calvo, Castrejón-Pita,
  Herrada \& Castrejón-Pita]{Ismail2018Softmatter}
{\sc \au{Ismail, A.~S.}, \au{Ga\~n\'an Calvo, A.~M.}, \au{Castrejón-Pita,
  J.~R.}, \au{Herrada, M.~A.} \& \au{Castrejón-Pita, A.~A.}} \yr{2018}
  \at{Controlled cavity collapse: scaling laws of drop formation}.  \jt{Soft
  Matter}  \bvol{14},  \pg{7671--7679}.

\bibitem[Ji {\em et~al.\/}(2021)Ji, Yang \& Feng]{ji2021compound}
{\sc \au{Ji, B.}, \au{Yang, Z.} \& \au{Feng, J.}} \yr{2021}  \at{Compound
  jetting from bubble bursting at an air-oil-water interface}.  \jt{Nature
  communications}  \bvol{12}~(1),  \pg{6305}.

\bibitem[Joung {\em et~al.\/}(2017)Joung, Ge \& Buie]{joung2017bioaerosol}
{\sc \au{Joung, Y.~S.}, \au{Ge, Z.} \& \au{Buie, C.~R.}} \yr{2017}
  \at{Bioaerosol generation by raindrops on soil}.  \jt{Nature Communications}
  \bvol{8},  \pg{14668}.

\bibitem[Kientzler {\em et~al.\/}(1954)Kientzler, Arons, Blanchard \&
  Woodcock]{kientzler1954photographic}
{\sc \au{Kientzler, C.~F.}, \au{Arons, Arnold~B.}, \au{Blanchard, D.~C.} \&
  \au{Woodcock, A.~H.}} \yr{1954}  \at{Photographic investigation of the
  projection of droplets by bubbles bursting at a water surface1}.  \jt{Tellus}
   \bvol{6}~(1),  \pg{1--7}.

\bibitem[Krishnan {\em et~al.\/}(2022)Krishnan, Bharadwaj \&
  Vasan]{krishnan2022impact}
{\sc \au{Krishnan, S.}, \au{Bharadwaj, S.~V.} \& \au{Vasan, V.}} \yr{2022}
  \at{Impact of freely falling liquid containers and subsequent jetting}.
  \jt{Experiments in Fluids}  \bvol{63}~(7),  \pg{1--20}.

\bibitem[Krishnan {\em et~al.\/}(2017)Krishnan, Hopfinger \&
  Puthenveettil]{krishnan_hopfinger_puthenveettil_2017}
{\sc \au{Krishnan, S.}, \au{Hopfinger, E.~J.} \& \au{Puthenveettil, B.~A.}}
  \yr{2017}  \at{On the scaling of jetting from bubble collapse at a liquid
  surface}.  \jt{Journal of Fluid Mechanics}  \bvol{822},  \pg{791-- 812}.

\bibitem[Krishnan \& Puthenveettil(2015)]{Sangeeth2015207}
{\sc \au{Krishnan, S.} \& \au{Puthenveettil, B.~A.}} \yr{2015}  \at{Dynamics of
  collapse of free surface bubbles}.  \jt{Procedia IUTAM}  \bvol{15},
  \pg{207--214}.

\bibitem[Krishnan {\em et~al.\/}(2020)Krishnan, Puthenveettil \&
  Hopfinger]{sangeeth_hole_expn}
{\sc \au{Krishnan, S.}, \au{Puthenveettil, B.~A.} \& \au{Hopfinger, E.~J.}}
  \yr{2020}  \at{Hole expansion from a bubble at a liquid surface}.
  \jt{Physics of Fluids}  \bvol{32}~(3),  \pg{032108},  \arxiv{arXiv:
  https://doi.org/10.1063/1.5139569}.

\bibitem[Lai {\em et~al.\/}(2018)Lai, Eggers \&
  Deike]{Universal_bubblejetcavity_Eggers_deike}
{\sc \au{Lai, C.~Y.}, \au{Eggers, J.} \& \au{Deike, L.}} \yr{2018}  \at{Bubble
  bursting: Universal cavity and jet profiles}.  \jt{Phys. Rev. Lett.}
  \bvol{121},  \pg{144501}.

\bibitem[Lee {\em et~al.\/}(2011)Lee, Weon, Park, Je, Fezzaa \&
  Lee]{san2011size}
{\sc \au{Lee, J.~S.}, \au{Weon, B.~M.}, \au{Park, S.~J.}, \au{Je, J.~H.},
  \au{Fezzaa, K.} \& \au{Lee, W.~K.}} \yr{2011}  \at{Size limits the formation
  of liquid jets during bubble bursting}.  \jt{Nature communications}
  \bvol{2},  \pg{367}.

\bibitem[Lighthill(1978)]{lighthill2001waves}
{\sc \au{Lighthill, J.}} \yr{1978} {\em Waves in fluids\/}.  \publ{Cambridge
  university press}.

\bibitem[MacIntyre(1972)]{MacIntyre1}
{\sc \au{MacIntyre, F.}} \yr{1972}  \at{Flow patterns in breaking bubbles}.
  \jt{Journal of Geophysical Research}  \bvol{77}~(27),  \pg{5211--5228}.

\bibitem[Oguz \& Prosperetti(1993)]{oguz1993dynamics}
{\sc \au{Oguz, H.~N.} \& \au{Prosperetti, A.}} \yr{1993}  \at{Dynamics of
  bubble growth and detachment from a needle}.  \jt{Journal of Fluid Mechanics}
   \bvol{257},  \pg{111--145}.

\bibitem[Perlin {\em et~al.\/}(1993)Perlin, Lin \& Ting]{perlin1993}
{\sc \au{Perlin, M.}, \au{Lin, H.} \& \au{Ting, C.~L.}} \yr{1993}  \at{On
  parasitic capillary waves generated by steep gravity waves: an experimental
  investigation with spatial and temporal measurements}.  \jt{Journal of Fluid
  Mechanics}  \bvol{255},  \pg{597--620}.

\bibitem[Puthenveettil {\em et~al.\/}(2018)Puthenveettil, Saha, Krishnan \&
  Hopfinger]{puthenveettil2018shape}
{\sc \au{Puthenveettil, B.~A.}, \au{Saha, A.}, \au{Krishnan, S.} \&
  \au{Hopfinger, E.~J.}} \yr{2018}  \at{Shape parameters of a floating bubble}.
   \jt{Physics of Fluids}  \bvol{30}~(11),  \pg{112105}.

\bibitem[Sampath {\em et~al.\/}(2019)Sampath, Afshar-Mohajer, Chandrala, Heo,
  Gilbert, Austin, Koehler \& Katz]{Sampath_GeoPhyRes2019}
{\sc \au{Sampath, K.}, \au{Afshar-Mohajer, N.}, \au{Chandrala, L.~D.}, \au{Heo,
  W.~S.}, \au{Gilbert, J.}, \au{Austin, D.}, \au{Koehler, K.} \& \au{Katz, J.}}
  \yr{2019}  \at{Aerosolization of crude oil-dispersant slicks due to bubble
  bursting}.  \jt{Journal of Geophysical Research: Atmospheres}
  \bvol{124}~(10),  \pg{5555--5578},  \arxiv{arXiv:
  https://agupubs.onlinelibrary.wiley.com/doi/pdf/10.1029/2018JD029338}.

\bibitem[Spiel(1995)]{spiel1995births}
{\sc \au{Spiel, D.~E.}} \yr{1995}  \at{On the births of jet drops from bubbles
  bursting on water surfaces}.  \jt{Journal of Geophysical Research: Oceans
  (1978--2012)}  \bvol{100}~(C3),  \pg{4995--5006}.

\bibitem[Thoroddsen {\em et~al.\/}(2018)Thoroddsen, Takehara, Nguyen \&
  Etoh]{thoroddsen_takehara_nguyen_etoh_2018}
{\sc \au{Thoroddsen, S.~T.}, \au{Takehara, K.}, \au{Nguyen, H.~D.} \& \au{Etoh,
  T.~G.}} \yr{2018}  \at{Singular jets during the collapse of drop-impact
  craters}.  \jt{Journal of Fluid Mechanics}  \bvol{848},  \pg{R3}.

\bibitem[Walls {\em et~al.\/}(2015)Walls, Henaux \& Bird]{Walls2015_drops}
{\sc \au{Walls, P. L.~L.}, \au{Henaux, L.} \& \au{Bird, J.~C.}} \yr{2015}
  \at{Jet drops from bursting bubbles: How gravity and viscosity couple to
  inhibit droplet production}.  \jt{Physical Review E}  \bvol{92},
  \pg{021002}.

\bibitem[Walls {\em et~al.\/}(2017)Walls, McRae, Natarajan, Johnson, Antoniou
  \& Bird]{walls2017quantifying}
{\sc \au{Walls, P. L.~L.}, \au{McRae, O.}, \au{Natarajan, V.}, \au{Johnson,
  C.}, \au{Antoniou, C.} \& \au{Bird, J.~C.}} \yr{2017}  \at{Quantifying the
  potential for bursting bubbles to damage suspended cells}.  \jt{Scientific
  reports}  \bvol{7}~(1),  \pg{15102}.

\bibitem[Woodcock {\em et~al.\/}(1953)Woodcock, Kientzler, Arons \&
  Blanchard]{woodcock1953giant}
{\sc \au{Woodcock, A.~H.}, \au{Kientzler, C.~F.}, \au{Arons, A.~B.} \&
  \au{Blanchard, D.~C.}} \yr{1953}  \at{Giant condensation nuclei from bursting
  bubbles}.  \jt{Nature}  \bvol{172},  \pg{1144--1145}.

\bibitem[Yang {\em et~al.\/}(2023)Yang, Ji, Ault \& Feng]{yang2023enhanced}
{\sc \au{Yang, Z.}, \au{Ji, B.}, \au{Ault, J.~T.} \& \au{Feng, J.}} \yr{2023}
  \at{Enhanced singular jet formation in oil-coated bubble bursting}.
  \jt{Nature Physics}  \pg{pp. 1--7}.

\bibitem[Yang {\em et~al.\/}(2020)Yang, Tian \&
  Thoroddsen]{yang_tian_thoroddsen_2020}
{\sc \au{Yang, Z.~Q.}, \au{Tian, Y.~S.} \& \au{Thoroddsen, S.~T.}} \yr{2020}
  \at{Multitude of dimple shapes can produce singular jets during the collapse
  of immiscible drop-impact craters}.  \jt{Journal of Fluid Mechanics}
  \bvol{904},  \pg{A19}.

\bibitem[Zeff {\em et~al.\/}(2000)Zeff, Kleber, Fineberg \&
  Lathrop]{zeff2000singularity}
{\sc \au{Zeff, B.~W.}, \au{Kleber, B.}, \au{Fineberg, J.} \& \au{Lathrop,
  D.~P.}} \yr{2000}  \at{Singularity dynamics in curvature collapse and jet
  eruption on a fluid surface}.  \jt{Nature}  \bvol{403}~(6768),
  \pg{401--404}.

\end{thebibliography}
\end{document}